\documentclass[preprint,a4paper,showabstract,superscriptaddress ]{revtex4-2}
\setcitestyle{super}
\usepackage{multirow}
\usepackage{bm}

\usepackage{graphicx}
\usepackage{epstopdf}
\usepackage{dcolumn}
\usepackage{bm}
\usepackage{hyperref}
\usepackage[utf8]{inputenc} 
\usepackage[T1]{fontenc}    
\usepackage{lmodern}
\usepackage{url}            
\usepackage{booktabs}       
\usepackage{amsfonts}       
\usepackage{nicefrac}       
\usepackage{microtype}      
\usepackage[version=3]{mhchem}
\usepackage{natbib}
\usepackage{amsmath}
\usepackage{amssymb}
\usepackage{xcolor}
\usepackage[range-units = single,range-phrase = \text{--},separate-uncertainty = true,detect-all]{siunitx}
	\DeclareSIUnit\linepair{lp}
	\DeclareSIUnit\pixels{px}
\usepackage{soul}
\setstcolor{red}
\usepackage{cancel}

\definecolor{bleudefrance}{rgb}{0.19, 0.55, 0.91}

\newcommand*{\corr}[1]{\textcolor{black}{#1}}

\newcommand{\inm}{\textrm{in}}
\newcommand{\outm}{\textrm{out}}

\newcommand{\xin}{x_\inm}
\newcommand{\xout}{x_\outm}

\newcommand{\rin}{\mathbf{r_\inm}}
\newcommand{\rout}{\mathbf{r_\outm}}
\newcommand{\rg}{\mathbf{r}}

\begin{document}

\title{Self-Portrait of the Focusing Process in Speckle: \\ II. Gouy Phase Shift for Defocus Correction \corr{and Pixel Depth Reassignment}}

\author{Flavien Bureau}
\affiliation{Institut Langevin, ESPCI Paris, PSL University, CNRS, 75005 Paris, France}
\author{Emma Brenner}
\affiliation{Institut Langevin, ESPCI Paris, PSL University, CNRS, 75005 Paris, France}
\author{Naiara Korta Martiartu}
\affiliation{Institut Langevin, ESPCI Paris, PSL University, CNRS, 75005 Paris, France}
\author{Elsa Giraudat}
\affiliation{Institut Langevin, ESPCI Paris, PSL University, CNRS, 75005 Paris, France}
\author{Arthur Le Ber}
\affiliation{Institut Langevin, ESPCI Paris, PSL University, CNRS, 75005 Paris, France}
\author{William Lambert}
\affiliation{SuperSonic Imagine, Aix-en-Provence, France}
\author{Louis Carmier}
\affiliation{Department of Radiology, Centre Hospitalier Universitaire de Montpellier, France}
\author{Aymeric Guibal}
\affiliation{Department of Radiology, Centre Hospitalier G\'{e}n\'{e}ral de Perpignan, France \\ \textnormal{{$^*$}Corresponding author (e-mail: alexandre.aubry@espci.fr)}}
\author{Mathias~Fink}
\affiliation{Institut Langevin, ESPCI Paris, PSL University, CNRS, 75005 Paris, France}
\author{Alexandre Aubry$^*$}
\affiliation{Institut Langevin, ESPCI Paris, PSL University, CNRS, 75005 Paris, France}

\date{\today}
\begin{abstract}
    \textbf{This is the second article in a series of three dealing with the exploitation of speckle for aberration correction and reverberation compensation in reflection imaging. When probing heterogeneous media with waves, we have to cope with multi-scale fluctuations of the wave velocity. On the one hand, short-scale heterogeneities induce back-scattered echoes whose random interference generate a speckle pattern on the beamformed image. On the other hand, large-scale fluctuations of the wave-velocity can distort the focused wave-fronts, resulting in aberrations on the same image. \corr{In this paper, we show how the self-portrait of the wave evolves as a function of the speed-of-sound model. Strikingly, a Gouy phase shift is observed when the speed-of-sound model is optimal. This particularly sensitive feature enables:} (\textit{i}) an optimization of the speed-of-sound model for each pixel of the image; (\textit{ii}) a local and fine compensation of defocus across the field-of-view, thereby compensating for most aberrations in the image. Experiment in a tissue-mimicking phantom and numerical simulations are first presented to validate our method. \corr{It is then applied to in-vivo liver data of a difficult-to-image patient. The speed-of-sound optimization allows an axial compensation of aberrations and a depth-reassignment of each singly-scattered echo to the actual position of the associated scatterer. As distance measurement is often critical for diagnosis, such a wave speed optimization can be crucial for ultrasound but also for any other imaging methods based on the principle of echo-location.}}
\end{abstract}
\maketitle
\corr{Classical reflection imaging methods, such as ultrasound, radar or optical coherence tomography, are based on the principle of echo-location. In acoustics, one or several transducers emit a wave toward the medium to be imaged. The incident wave is reflected by the heterogeneities and the backscattered wave field is measured by the same sensor(s). Under a single scattering assumption, each recorded echo is the result of a scattering event on each heterogeneity. Under a constant wave velocity assumption, the time-of-flight $t$ of each echo indicates the distance $r$ at which lies the associated reflector, such that $r=c_0t/2$. To discriminate echoes coming form scatterers located at a same distance $r$, a focusing process is nevertheless required. Physically, this can be done by emitting and scanning a focused incident wave-field. Numerically, this can be done by applying a delay-and-sum beamforming process to the recorded echoes. Signals from a particular echo are selected by computing the time-of-flight
associated with the forward and return travel paths of the ultrasonic wave between each transducer and the image voxel. From a physical point of view, time delays in transmission are used to concentrate the ultrasound wave on a focal area and time delays at reception select echoes coming from this excited area. The critical step of computing the time-of-flight for each insonification and each focal point is achieved in any clinical
device by assuming the medium as homogeneous with a constant speed of sound. This assumption is necessary in order to achieve the rapidity required for real-time imaging; however, it may not be valid for some configurations in which long-scale fluctuations of the speed-of-sound impact wave propagation~\cite{hinkelman_measurements_1997}.}

\corr{The mismatch between the wave velocity model and speed-of-sound distribution  results in an axial shift between each focal spot and the isochronous volume, the ensemble of points that contribute to the detected signal at a given time-of-flight $t$. This defocus results in a loss of resolution and contrast that fluctuates over the image. Speed-of-sound fluctuations can also generate longitudinal distortions of the medium reflectivity on the image. The axial dimension is actually dictated by the time-of-flight $t$ of echoes. The obtained image is therefore not representative of the true depth of scatterers inside the medium. Those problems are particularly frequent in soft tissues since the speed of sound typically ranges from 1400 m/s (e.g. fat tissues) to 1650 m/s (e.g. skin, muscle tissues)~\cite{duck_acoustic_1990}. Moreover, many diagnoses comprise distance measurements based on the ultrasound image~\cite{Scorza2015}. In that regard, a depth-reassignmement of each echo would be particularly relevant.}

\corr{To tackle aberrations, adaptive focusing techniques~\cite{hirama_adaptive_1982,odonnell_phase-aberration_1988,nock_phase_1989,mallart_adaptive_1994,masoy_iteration_2005,dahl_phase_2005,montaldo_time_2011} inspired from astronomy~\cite{babcock_possibility_1953,labeyrie_attainment_1970} have been developed since the eighties. However, they required a tedious and iterative focusing process that made them illusory for real-time imaging. Recently, the drastic gain in computational power and memory has given a new birth to the field. Based on the experimental measurement of the reflection matrix, adaptive focusing strategies can now be performed in post-processing~\cite{chau_locally_2019,lambert_distortion_2020,bendjador_svd_2020,bureau_three-dimensional_2023}. Nevertheless, the designed focusing laws tend to shift each focal spot towards the distorted isochronous volume. Hence axial aberrations subsist and the depth position of scatters remains uncertain.}

\corr{To circumvent this issue, an alternative approach is to map the speed-of-sound and use it to design an adapted beamforming process. This is the principle of computed ultrasound tomography in echo mode (CUTE)~\cite{jaeger_full_2015,Staehli2023}. This powerful approach shows nevertheless a bias with respect to the initial speed-of-sound model~\cite{stahli_improved_2020}. Less ambitious but more robust with respect to the latter issue, other approaches consist in determining the speed-of-sound that optimizes the focusing process at each point of the medium~\cite{imbault_robust_2017,jakovljevic_local_2018,ali_distributed_2022}. This optimal wave velocity is not the local speed-of-sound but the inverse of the mean slowness integrated between the probe and this  point. In previous studies, the focusing process was optimized by maximizing focusing parameters such as the coherence factor~\cite{jakovljevic_local_2018,ali_distributed_2022,Vraalstad2024}, the image brightness~\cite{benjamin_novel_2018,zubajlo_experimental_2018,napolitano_sound_2006,cho_efficient_2009,yoon_optimal_2012,yoon_vitro_2011,shin_estimation_2010,perrot_so_2021} or a focusing criterion~\cite{lambert_ultrasound_2022a} based on the reflection point spread function~\cite{lambert_reflection_2020}. In the present paper, we will exploit the self-portrait of the coherent wave~\cite{Giraudat2025} and its dependence with respect to the wave velocity model in order to map the optimal speed-of-sound over the field-of-view.}

\corr{Compared to previous studies, a wave speed optimization based on the coherent wave-field has two main advantages: (i) the prior filtering of multiple scattering and noise that usually pollute in-vivo ultrasound data~\cite{lambert_ultrasound_2022a}; (ii) the Gouy phase shift~\cite{feng_physical_2001} exhibited by this wave-field upon focusing. We will show how the latter phenomenon can lead to a sharper estimation of the optimal speed-of-sound. Beyond improving the contrast and resolution of a standard ultrasound image in clinics~\cite{Vraalstad2024}, the knowledge of an optimized speed-of-sound is also extremely rewarding for an axial compensation of aberrations. The axial dimension of the ultrasound image is no longer dictated by the echo time but can be rescaled as a function of depth~\cite{Ahmed2024}. As said previously, the reassignment of each scatterer to its true depth is anything but trivial, since numerous diagnosis protocols are based on distance measurement.} 

\corr{To validate our approach, we will first consider a tissue-mimicking phantom. We will show how the coherent component of the wave-field and its dependence with respect to the wave speed can be extracted in speckle. Mathematically, this will be done by a singular value decomposition of the de-scan reflection matrix~\cite{Giraudat2025} parameterized as a function of the wave speed model. Then, we will validate and outline the limits of our method by considering numerical experiments in synthetic samples in which the speed-of-sound distribution is known. At last, to demonstrate its potential for medical imaging, we will apply it to speed-of-sound measurement in the liver for a difficult-to-image patient suffering from hepatic steatosis. 
This disease consists in an accumulation of fat droplets that results in a low speed-of-sound ($c\sim 1540$ m.s$^{-1}$) compared to its usual value in liver  ($c\sim 1585$ m.s$^{-1}$)~\cite{Staehli2023}. The effectiveness of ultrasound for diagnosing hepatic steatosis is reduced in obese patients~\cite{Almeida2008}. Indeed, because the ultrasonic waves must travel through successive layers of skin, fat, and muscle tissue before reaching the liver, both the incident and reflected wave-fronts undergo strong aberrations~\cite{Hinkelman1998,Browne2005} and reverberations (clutter noise)~\citep{Lediju2009}. 
Despite an ultrasound image degraded by those detrimental phenomena, our matrix approach will compensate for axial aberrations and highlight coherent variations of the optimal speed-of-sound across fat, muscles and liver. A particularly low value will be found for the sound speed in liver, which is in agreement with the steatosis diagnosis. In the last part of the paper, we will discuss the merits and limits of our approach as well as its potential combination  with more sophisticated methods such as CUTE~\cite{jaeger_computed_2015,Staehli2023,heriard-dubreuil_refraction-based_2023} or speed-of-sound optimization approaches~\cite{Simson2024}.}

\section{Focused reflection matrix} 

\subsection{Reflection matrix {acquisition}} 

\begin{figure*}[t!]
  \includegraphics[width=1\linewidth]{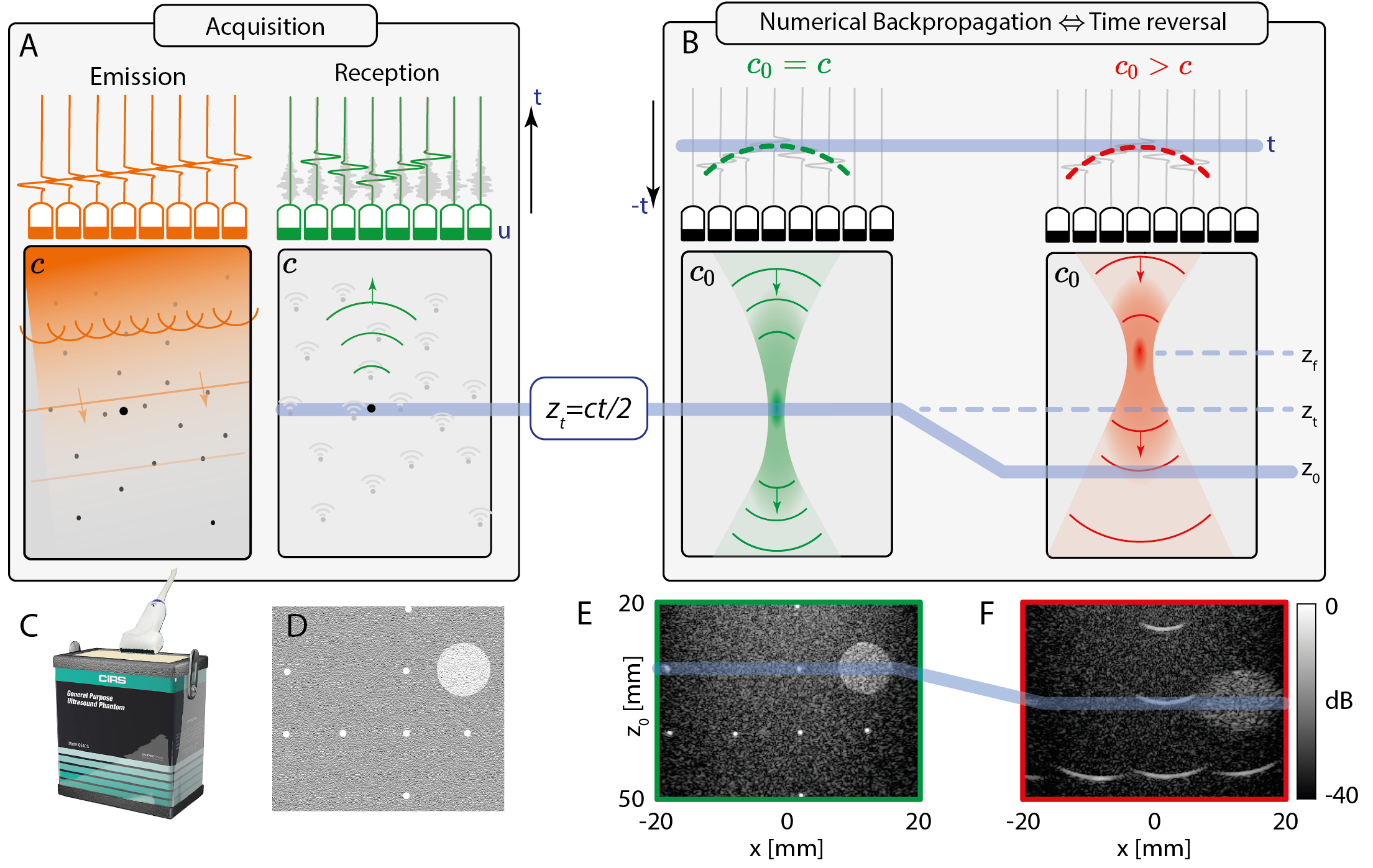}
    \caption{\textbf{{Impact of {an incorrect} speed-of-sound model in ultrasound imaging.}} {(\textit{A})} The acquisition of the reflection matrix consists in insonifying the medium with a set of plane waves emitted by the ultrasonic probe. The recorded wave-fronts are stored in the reflection matrix. The contribution of one scatterer located at depth $z_t$ is highlighted in green. For sake of simplicity, the wave velocity $c$ is considered as homogeneous.  {(\textit{B})} The numerical focusing process can be seen as a fictive time reversal experiment in a medium of wave velocity $c_0$. If $c_0=c$, the time-reversed wave-front back-focuses exactly at the initial scatterer location. If $c_0\neq c$, a mismatch exists between the focusing plane ($z_f=cz_t/c_0$), the isochronous plane ($z_t=c t/2$) and the imaging plane  ($z_0=c_0 t/2$). {(\textit{C})} Experimental configuration: A linear array of transducers is placed on top of an ultrasound phantom. (\textit{D}) Scheme of the phantom with nylon rods (white), random distribution of unresolved scatterers (gray) and a more echogene cylinder displaying a stronger concentration of scatterers (light gray). (\textit{E{, F}}) Corresponding ultrasound image for $c = c_0 = 1540 $ m.s$^{-1}$ and $c\neq c_0=1800$ m.s$^{-1}$, respectively.}    
    \label{fig_UMI}
\end{figure*}

\corr{Ultrasound matrix imaging (UMI) begins with the acquisition of the reflection matrix. A linear array of transducers ({SL15-4}, Supersonic Imagine) whose characteristics are provided in Tab.~\ref{AcquisitionParam} is placed in direct contact of the medium (Figs.~\ref{fig_UMI}A,B). The probe is controlled by a medical ultrafast ultrasound scanner (Aixplorer Mach-30, Supersonic Imagine, Aix-en-Provence, France). The first sample under study is a tissue mimicking phantom with a speed-of-sound $c=1542 \pm 10$ m/s (Fig.~\ref{fig_UMI}C). It is composed of a random distribution of unresolved scatterers which generate an ultrasonic speckle characteristic of human tissue (Fig.~\ref{fig_UMI}D). The reflection matrix is captured by sending a series of plane waves into the medium (Fig.~\ref{fig_UMI}A). The parameters of this emission sequence are given in Tab.~\ref{AcquisitionParam}. Plane waves are generated assuming a constant speed of sound in the medium $c_{\textrm{acq}}=1540$ m/s. Each plane wave is identified with its angle of incidence ${\theta}_{\textrm{in}}$.} For each illumination, the reflected waves are recorded by the transducers of the probe, each element being identified by its lateral position $u_\outm$ (Fig.~\ref{fig_UMI}B). The recorded wave-fronts are noted $R(u_\outm,{\theta}_{\textrm{in}},t)$, with $t$ the echo time. They are stored in a reflection matrix ${\mathbf{R}_{\mathbf{u}\boldsymbol{\theta}}(t)=[R(u_\outm,\theta_{\textrm{in}},t)]}$.
\begin{table}[!ht]
    \center
\begin{tabular}{l|lll}
      \multicolumn{2}{c}{\textbf{}} & \multicolumn{1}{l}{\textbf{Phantom}} & \multicolumn{1}{l}{\textbf{Liver}} \\
      \hline
    \multicolumn{2}{l}{{Probe type}}  & Linear & Curve \\
    \multicolumn{2}{l}{{Curvature radius} $\mathcal{R}_u$}  & / & $60$  mm\\
    \multicolumn{2}{l}{Number of transducers $N_{u_\outm}$} & $128$ & 192 \\ 
    \multicolumn{2}{l}{Transducer pitch $\delta u/\delta \Theta_{u}$}  & $0.2$ mm & $0.32^\circ$ \\
    \multicolumn{2}{l}{{Central frequency} $f_\textrm{c}$} & $7.5$ MHz & $3.5$ MHz \\
    \multicolumn{2}{l}{{Bandwidth} $ \Delta f$} & ${[4-15]}$ MHz & $[1-6]$ MHz\\
    \multirow{3}{*}{{Plane waves}}
        & Maximum $\theta_\inm^{(\textrm{max})}$& $40^\circ$ & $20^\circ$ \\ 
        & Pitch $\delta \theta_\inm$& $1^\circ$ & $1^\circ$ \\ 
        & Number $N_{\theta_\inm}$ & $81$ & $41$\\
    \multicolumn{2}{l}{{Sampling frequency} $f_\textrm{s}$} & $30$ MHz& $26.7$ MHz\\ 
    \multicolumn{2}{l}{{Recording time} $\Delta t$} & $137$ $\mu$s&  $235$ $\mu$s\\
    \hline
\end{tabular}
\caption{{Acquisition parameters in the phantom and liver experiments.} }
\label{AcquisitionParam}
\end{table}

\subsection{Confocal imaging} 
The first post-processing step is to build a confocal image $\mathcal{I}$ from the recorded reflection matrix. To do so, a delay-and-sum beamforming process is applied to the coefficients of ${\mathbf{R}_{\mathbf{u}\boldsymbol{\theta}}}(t)$. Physically equivalent to a confocal focusing process (Fig.~\ref{fig_UMI}B), this procedure writes mathematically as follows:
\begin{align}
\label{confocal}
\mathcal{I}(x,z_0=c_0t/2)&=  \sum_{\theta_\inm}\sum_{u_\outm} A(u_\outm,\theta_\textrm{in},x,t,c_0)  \\  & R(u_\outm,\theta_\textrm{in},\tau_{\textrm{out}}(u_\outm,x,t,c_0)
+\tau_{\textrm{in}}(\theta_\textrm{in},x,t,c_0) \corr{+\tau_L}). \nonumber
\end{align}
$c_0$ is the wave velocity model considered in the beamforming process. $\tau_{\textrm{in}}$ is the time-of-flight expected for the incident plane wave to reach the target point of coordinates $(x,z_0)$ (Appendix~\ref{K}). $\tau_{\textrm{out}}$ is the time-of-flight expected for the reflected wave to travel from the same target point to each transducer (Appendix~\ref{K}). \corr{$\tau_L$ is a time shift that accounts for the acoustic lens that sits at the trasducer array surface in order to collimate the ultrasonic beams in a 2D plane}. $A$ is a normalization and apodization factor that limits the extent of the
receive synthetic aperture. $z_0=c_0t/2$ is the expected position of the isochronous volume, which is defined as the ensemble of points that contribute to the ultrasound signal at time $t$. If the wave velocity model is correct ($c_0=c$), $z_0$ is a relevant estimator of the scatterers' depth and the ultrasound image is a satisfactory image of the medium reflectivity (Fig.~\ref{fig_UMI}E). On the contrary, if the wave velocity model is incorrect ($c_0\neq c$), each detected scatterer at depth $z_t=ct/2$ is assigned to a false depth $z_0$ (Fig.~\ref{fig_UMI}F): \corr{$z_0=(c_0/c)z_t $}. Moreover, the beamformed image is drastically affected by the mismatch between the isochronous volume at $z_t$ and the focusing plane at $z_f$ (Fig.~\ref{fig_UMI}B) since  $z_f=(c/c_0)z_t$ (Appendix~\ref{B}). This non-coincidence result in axial aberrations on the ultrasound image (Fig.~\ref{fig_UMI}F) that manifest as: (\textit{i}) an axial shift of the scatterers with respect to their true depth; (\textit{ii}) a degradation of the transverse resolution. The search for the optimum speed-of-sound $c$ for a particular point therefore consists of bringing the imaging plane into coincidence with the focal plane~\cite{perrot_so_2021}. If such an optimization seems trivial when considering a bright spot, the goal is now to develop a method to find this value in random speckle. Indeed, the speckle statistics seems unaffected by the wave velocity used in the beamforming process (Figs.~\ref{fig_UMI}E and F). 

\subsection{Focused reflection matrix} 

UMI can provide a solution to this fundamental issue. The focusing quality can be assessed locally in the ultrasound speckle by projecting the reflection matrix in a focused basis~\cite{lambert_reflection_2020}. In the time domain, this operation can be performed by decoupling the input and output focal spots in the beamforming process~\cite{lambert_ultrasound_2022a}:
\begin{align}
\label{focusedR}
 R &(x_\textrm{out},x_\textrm{in},t,c_0)=\sum_{\theta_\inm}\sum_{u_\outm}  A(u_\outm,\theta_\textrm{in},x_\textrm{out},t,c_0)  \\ & R(u_\outm,\theta_\textrm{in}, \tau_\textrm{out}(u_\outm,x_\textrm{out},t,c_0)
+\tau_\textrm{in}(\theta_\textrm{in},x_\textrm{in},t,c_0)\corr{+\tau_L}),\nonumber
\end{align}
At each echo time $t$, the focused reflection matrix $\mathbf{R}_{xx}(t,c_0)$ contains the response $R(x_\textrm{out},x_\textrm{in},t,c_0)$ between virtual transducers at $\rin=(x_\textrm{in},t)$ and $\rout=(x_\textrm{out},t)$ (Fig.~\ref{fig_rpsf}A) whose axial positions corresponds to the depth $z_t$ of the isochronous volume and thus dictated by the time-of-flight $t$. The diagonal elements of each matrix $\mathbf{R}_{xx}(t,c_0)$ considered at the ballistic time $t=2z_0/c_0$ directly correspond to the confocal image that we previously introduced. However, $\mathbf{R}_{xx}(t,c_0)$ contains much more information than the confocal image: The spreading of energy over its off-diagonal elements is an indicator of the focusing quality in speckle by probing the cross-talk between distinct virtual transducers~ \cite{lambert_reflection_2020,lambert_ultrasound_2022a}. \corr{In contrast with the other papers of the series~\cite{Giraudat2025,Giraudat2025b}, we are here not interested in the problem of reverberations. Instead, we want to evaluate the focusing process of the ballistic component with respect to the wave velocity model. This is why we restrict our study of the focused reflection matrix to the ballistic time and to virtual transducers placed at the same depth.}

\section{Self-portrait of the focusing process}

\subsection{De-scan reflection matrix}
\begin{figure}[htbp]
\includegraphics[width=14cm]{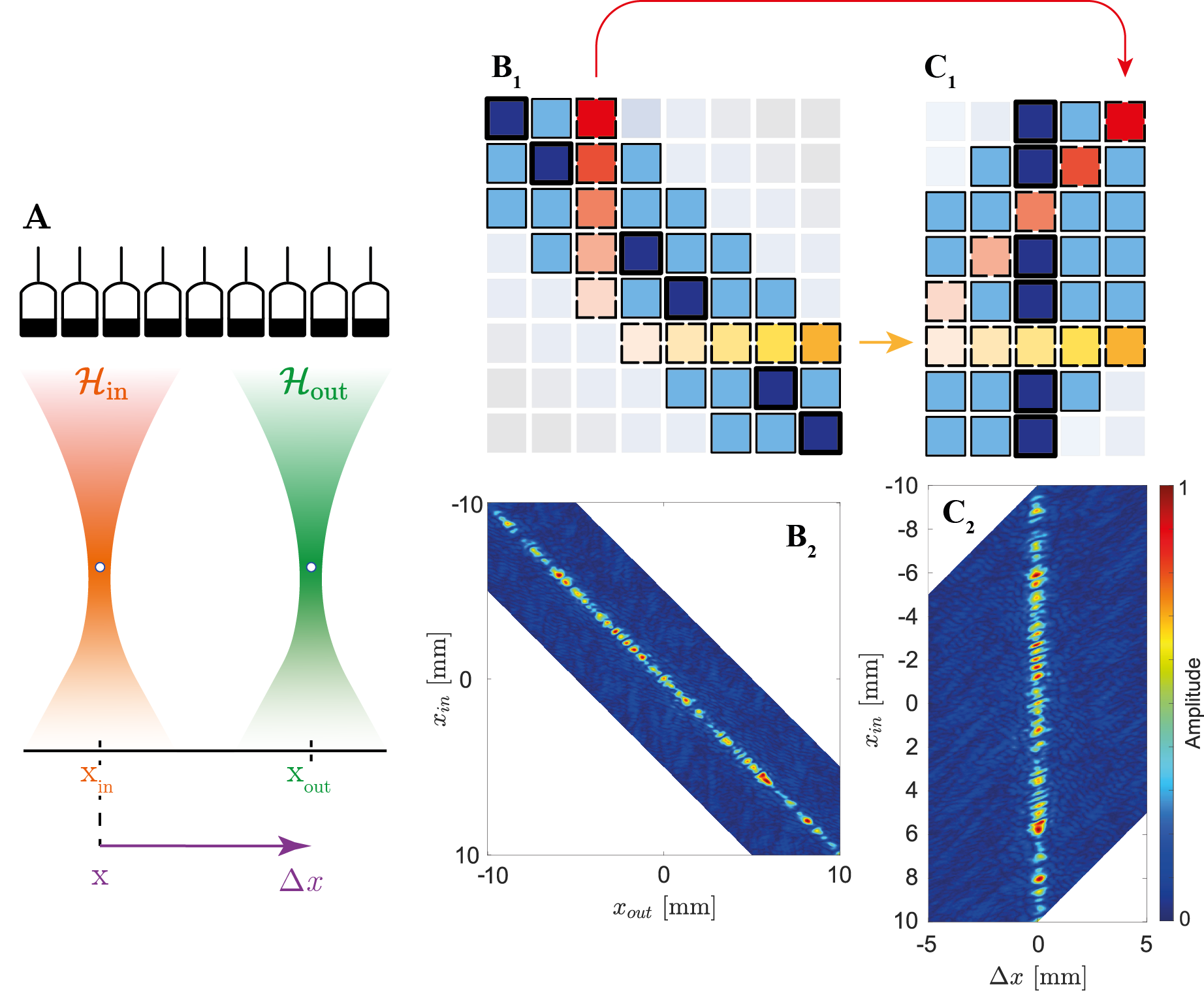}   \caption{\textbf{The de-scan focused basis.} ({A}) Schematic view of the input and output focal spots. ({B}) 
Reflection matrix $\mathbf{R}=[R(\xin,\xout,t,c_0)]$ expressed in the conventional focused basis. ({C}) Reflection matrix $\mathbf{R}_{\mathcal{D}}=[R(x_{\textrm{in}},t,\Delta x,c_0)]$ expressed in the de-scanned basis, with $\Delta x=\xout-\xin$. The sub-panels {B}$_2$ and {C}$_2$ are examples of reflection matrices sketched in sub-panels \textbf{B}$_1$ and {C}$_1$, respectively. They correspond to the tissue mimicking phantom experiment (Table~1) for time $t=32\mu$s and speed-of-sound $c_0=1540$ m.s$^{-1}$. }
    \label{figsupp_descan}
\end{figure}

To investigate this cross-talk, the focused reflection matrix can be expressed in a de-scanned frame \corr{(Fig.~\ref{figsupp_descan})}. Mathematically, it consists in re-arranging the ultrasound data using the following change of variables:
\begin{equation}
R_D(\boldsymbol{\Delta},\mathbf{r}_\textrm{in} )=R( x_\textrm{out}, x_\textrm{in},  t,c_0)
\end{equation}
with $\boldsymbol{\Delta} =(\Delta x,c_0) $ and $\Delta x=x_\textrm{out}-x_\textrm{in}$, the relative lateral distance between virtual transducers. Each column of the resulting de-scan matrix $\mathbf{R}_D$ shows the reflected wave-field in the imaging plane re-centered around each input focusing point $\mathbf{r}_\textrm{in}$. We will refer to this quantity as the reflection point spread function (RPSF)~\cite{lambert_reflection_2020}. Each RPSF is identified by its transverse position $x_\textrm{in}$ and echo time $t$. The lateral extension $\Delta x$ of the RPSF is investigated as a function of the wave velocity model $c_0$.

Figure~\ref{fig_rpsf}B shows three realizations of RPSFs obtained for different speckle grains $\mathbf{r}_\textrm{in}$ displayed in Figure~\ref{fig_rpsf}{C}. These RPSFs display a focal spot whose spatial extension is minimized for $c=c_0$. However, each RPSF is modulated by the random reflectivity of the sample. To get rid of this problem, the solution is to perform a local average of the focal spots in order to unscramble the effect of wave propagation from the sample reflectivity. To that aim, the field of view shall be truncated into overlapping spatial windows $\mathcal{P}(\rin-\rg_\textrm{p})$ defined by their center $\rg_\textrm{p}$ and their spatio-temporal extent $\mathbf{p}=(p_x,p_t)$, where $p_x$ and $p_t$ denote the lateral and axial extent of each window, respectively. A local reflection matrix $\mathbf{R}_\textrm{L}(\rg_\textrm{p})$ can then be defined for each point $\rg_\textrm{p}=(x_p,t_p)$ in the field-of-view. Its coefficients write 
\begin{align}
\label{P}
    R_\textrm{L}(\boldsymbol{\Delta},\rg,\rg_\textrm{p})=R_D(\boldsymbol{\Delta},\rg)\mathcal{P}(\rg-\rg_\textrm{p}),
\end{align}
with $\mathcal{P}(\rg-\rg_\textrm{p})=1$ for $\lvert x-x_\textrm{p}\rvert <p_x /2$ and $\lvert t-t_\textrm{p}\rvert <p_t /2$, and zero otherwise.

\subsection{Incoherent {Reflection} Point Spread Function}

\begin{figure*}
\centering
\includegraphics[width=1\linewidth]{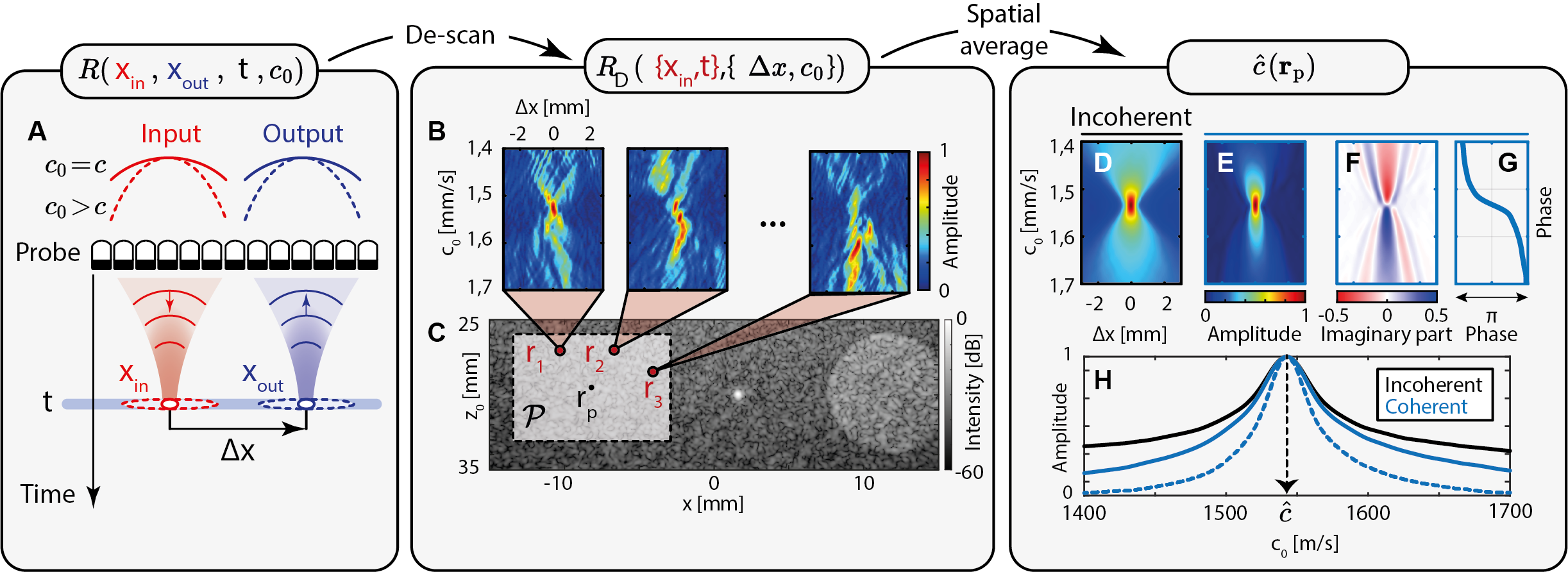}
    \caption{\textbf{Self-portrait of the focusing process.} ($A$) Matrix imaging consists in splitting the input and the output focusing points during the beamforming process. The focused reflection matrix allows the monitoring of the focusing process with respect to the wave velocity model $c_0$. ($B$) Such a matrix can be expressed in a de-scanned basis in order to provide the dependence of the RPSF shown here in amplitude for three speckle spots of the ultrasound image. ($C$) The three speckle grains considered in panel B and the area $\mathcal{P}$ considered for the local averaging of the RPSF in panels D-H are superimposed to the ultrasound image of the phantom. ($D$) Incoherent RPSF. ($E$) Amplitude of the coherent RPSF. ($F$) {Imaginary part} of the coherent RPSF. ($G$) Phase of the coherent RPSF as a function of $c_0$ at $\Delta x=0$. ($H$) {Magnitude of the incoherent RPSF (black line), of the coherent RPSF (blue line) and its real part (dashed blue) as a function of $c_0$ at $\Delta x=0$.}  Spatial averaging is here performed with a window of size $(p_x,p_t)=(10$ mm, $1.3$ $\mu$s) centered around $(x,t)=(0$ mm, $43$ $\mu$s).}
    \label{fig_rpsf}
\end{figure*}

The most direct way for probing the focusing quality is to perform a local and incoherent average of each RPSF:
\begin{equation}
        RPSF_{\textrm{inc}}(\boldsymbol{{\Delta}},\rg_\textrm{p})=\sqrt{\left\langle \left\lvert R_{L}(\boldsymbol{{\Delta}},\mathbf{r},\rg_\textrm{p} )\right\lvert^2\right\rangle_{\mathbf{r}}},
        \label{RPSFdefsos}
\end{equation}
where $\langle \cdots \rangle$ denotes a spatial average over the different speckle grains $\mathbf{r}$. The result is displayed in Figure~\ref{fig_rpsf}{D} for the area $\mathcal{P}$ indicated in Fig.~\ref{fig_rpsf}B. As expected, spatial averaging tends to smooth out reflectivity fluctuations. More quantitatively, its intensity provides, in the speckle regime, an estimation of the auto-convolution of the transmit and receive PSFs, $h_\textrm{in}$ and $h_\textrm{out}$, respectively~\cite{lambert_reflection_2020} (see Appendix~\ref{D}):  \begin{equation}
RPSF^2_{\textrm{inc}}(\Delta x,c_0,\rg_\textrm{p} ) \propto |h_\textrm{in}|^2 \stackrel{\Delta x}{\circledast} |h_\textrm{out}|^2 (\Delta x,c_0,\rg_\textrm{p}).
\end{equation} 
A direct estimation of the speed-of-sound can be obtained by considering the value of $c_0$ that maximizes the amplitude of this incoherent RPSF (black line in Fig. \ref{fig_rpsf}H):
\begin{equation}
    \hat{c}_{\textrm{inc}}(\rg_{\textrm{p}})=\underset{\boldsymbol{\Delta}}{\textrm{argmax}} \left ( RPSF_{\textrm{inc}} ( \boldsymbol{\Delta},\rg_\textrm{p}) \right).
    \label{cinc}
\end{equation}
The estimated speed-of-sound is $\hat{c}_\textrm{inc}=1543$ m/s, which is within the uncertainty margin provided by the manufacturer. Nevertheless, the incoherent {RPSF} also displays a strong background induced by multiple scattering events that can hamper the estimation of the speed-of-sound in more complex situations. Under a Gaussian beam approximation, the uncertainty of this measurement can actually be expressed as follows (Appendix~\ref{H}): 
\begin{equation}
\label{deltacinc}
\delta c_{\textrm{inc}}  = \frac{2}{\sqrt{3}}\frac{1}{\beta^{1/2}{N_{\mathcal{P}}}^{1/4} }\frac{z_R}{t}
\end{equation}
with $\beta$, the signal-to-noise ratio, $z_R \sim 2\lambda/NA^2$, the depth-of-field (or Rayleigh range), $NA$, the numerical aperture and $N_{\mathcal{P}}$, the number of independent speckle grains in each area $\mathcal{P}$. 
This last equation points out the main parameters that control the error of our wave speed estimator. Not surprisingly, $\delta c_{\textrm{inc}}$ is directly impacted by the signal-to-noise ratio and decreases as $\beta^{-1/2}$. An inverse scaling is observed with the time-of-flight $t$, which reflects the fact that the precision on the speed-of-sound measurement improves with the travel path length. Interestingly, the linear dependence of $\delta c_{\textrm{inc}}$ with $z_R$ implies a sharper measurement at high numerical aperture {(Appendix~\ref{I})}. Through the scaling of $\delta c_{\textrm{inc}} $ as $N_{\mathcal{P}}^{-1/4}$, Eq.~\ref{deltacinc} also highlights the compromise we will have to make further for wave velocity tomography. On the one hand, each patch $\mathcal{P}$ should encompass a sufficient number of resolution cells in order to reduce the bias of the speed-of-sound estimator. On the other hand, the size of each patch will control the spatial resolution of the sound speed map.  

\subsection{Revealing the coherent wave}

To reduce the uncertainty, a coherent RPSF can be extracted through a singular value decomposition of the local matrix $\mathbf{R}_{L}(\rg_\textrm{p})$:
\begin{align}
\label{SVD}
    \mathbf{R}_L(\rg_\textrm{p})= \mathbf{U} \times \mathbf{\Sigma} \times \mathbf{V}^\dagger
\end{align}
where $\mathbf{\Sigma}$ is a diagonal matrix containing the singular value{s} $\sigma_i$ in descending order: $\sigma_1>\sigma_2>..>\sigma_N$. $\mathbf{U}$ and $\mathbf{V}$ are unitary matrices that contain the orthonormal set of eigenvectors, $\mathbf{U}_i=[U_i( \Delta x,c_0  )]$ and $\mathbf{V}_i=[V_i(\mathbf{r})]$. In first approximation, the de-scanned matrix is of rank 1 (Appendix~\ref{E}). The first singular vector $\mathbf{U}_1$, directly provides a coherent RPSF, which is a direct estimator of the output PSF weighted by the confocal value of the input PSF (Appendix~\ref{E}): 
\begin{equation}
\label{PSFcoh}
RPSF_{\textrm{coh}} (\Delta x, c_0,\rg_\textrm{p})   \propto {h}_{\textrm{in}}(0,c_0,\rg_\textrm{p}) {h}_{\textrm{out}}(\Delta x,c_0,\rg_\textrm{p}).
\end{equation}
The amplitude of the coherent {RPSF} obtained in the ultrasound phantom is displayed in Fig.~\ref{fig_rpsf}E for the area $\mathcal{P}$ indicated in Fig.~\ref{fig_rpsf}B. Compared to its incoherent counterpart (Fig.~\ref{fig_rpsf}D), the multiple scattering background has been reduced, which provides a more contrasted view of the focusing quality in the phantom (Fig.~\ref{fig_rpsf}H). A novel estimation of the speed-of-sound can be performed by probing its maximum value (blue curve in Fig.~\ref{fig_rpsf}H):
\begin{equation}
    \hat{c}_{\textrm{coh}}(\rg_{\textrm{p}})=\underset{\boldsymbol{\Delta}}{\textrm{argmax}} \left ( |RPSF_{\textrm{coh}}| ( \boldsymbol{\Delta},\rg_\textrm{p}) \right);
    \label{ccoh}
\end{equation}
The uncertainty $\delta c_{\textrm{coh}}$ of such a measurement is slightly better than its incoherent counterpart (Eq.~\ref{deltacinc}) since \corr{$\delta c_{\textrm{coh}}={\sqrt{3}}\delta c_{\textrm{inc}}/{2}$} (Appendix~\ref{H}). 

\subsection{Exploiting the Gouy phase}
Interestingly, the uncertainty of Eq.~\ref{deltacinc} can be again reduced by leveraging the phase of the coherent RPSF (Fig.~\ref{fig_rpsf}F). A phase jump is actually observed in the vicinity of the optimal wave speed (Fig.~\ref{fig_rpsf}G). This feature is equivalent to the Gouy phase shift $\phi_G$ generally exhibited by a focused wave in the focal plane~\cite{feng_physical_2001}. The originality of our observation here is that it occurs when the model speed-of-sound coincides with the sound velocity of the phantom. While $\phi_G$ should be of $\pi/2$ in a 2D configuration, the phase of the coherent RPSF (Fig.~\ref{fig_rpsf}{G}) shows a shift of 2$\phi_G=\pi$ due to the confocal nature of the measured RPSF (Eq.~\ref{PSFcoh}). \corr{The} Gouy phase shift originates from the transverse spatial confinement of the wave-field and can be thus a relevant observable for speed-of-sound estimation.

The information carried by the phase of the coherent RPSF can be exploited by considering the real part of the coherent RPSF, $\mathcal{R} \left [RPSF_{\textrm{coh}} \right ]$. The maximization of this quantity leads to a new estimator $\hat{c}_\textrm{gouy}$ of the sound speed: 
\begin{equation}
    \hat{c}_{\textrm{gouy}}(\rg_{\textrm{p}})=\underset{\boldsymbol{\Delta}}{\textrm{argmax}} \left ( \mathcal{R} \left [RPSF_{\textrm{coh}} ( \boldsymbol{\Delta},\rg_\textrm{p}) \right ]\right).
    \label{cgouy}
\end{equation}
$\hat{c}_\textrm{gouy}$ actually exploits the amplitude enhancement and the phase jump of the RPSF to provide a sharper estimation of the speed-of-sound by almost a factor 3 with respect to $\hat{c}_\textrm{inc}$ (Appendix~\ref{H}): $\delta {c}_\textrm{gouy}=\delta {c}_\textrm{coh}/\sqrt{5}$. This better precision is highlighted by the steeper peak centered around $\hat{c}_{\textrm{gouy}}={1542.5}$ m/s exhibited by the real part of the RPSF in Fig.~\ref{fig_rpsf}H. 

\section{Axial compensation of aberrations}

\subsection{Local compensation of defocus}
\begin{figure*}[htbp]
\centering
\includegraphics[width=16cm]{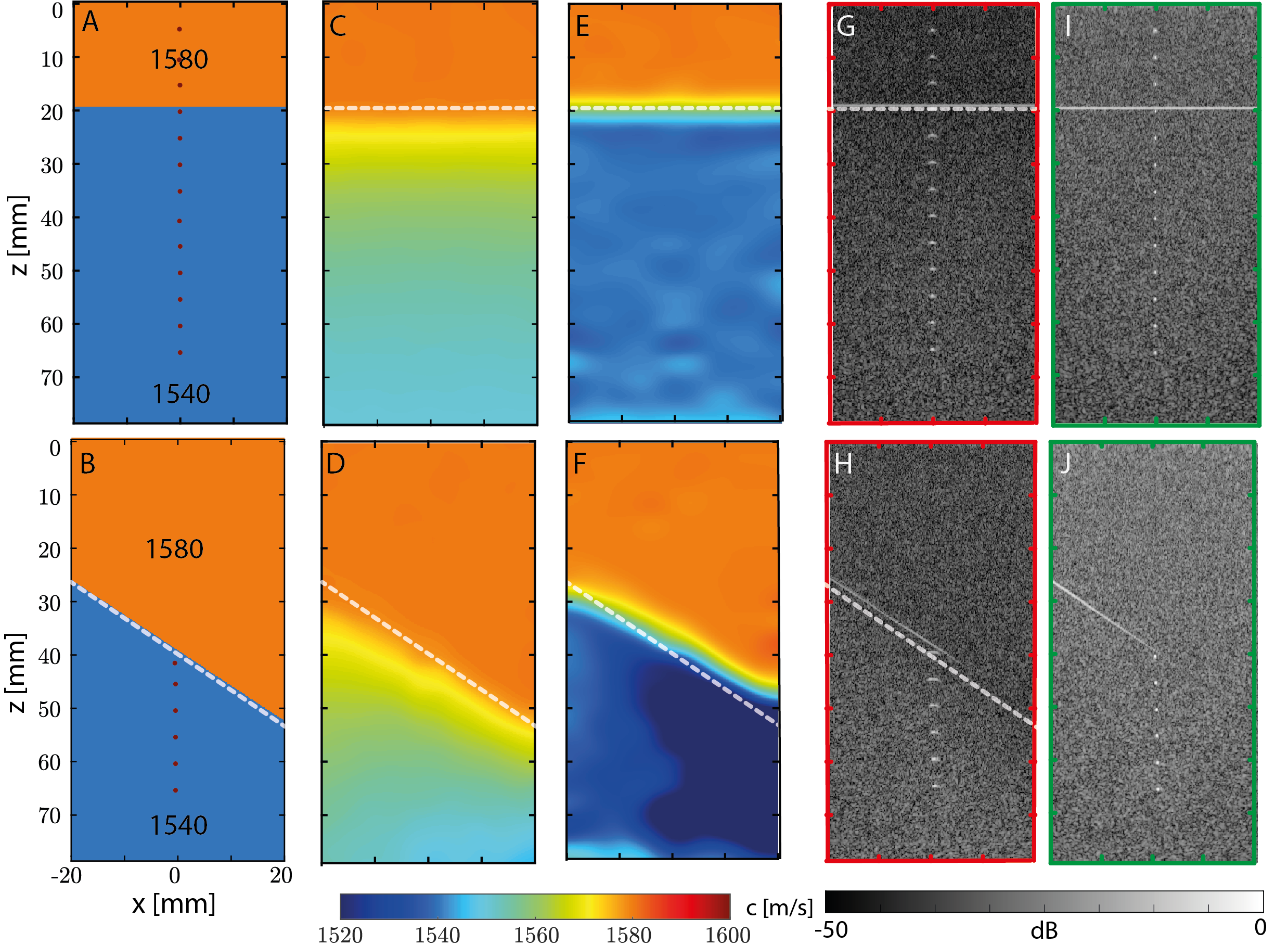}
    \caption{\textbf{Numerical validation of defocus compensation.} (${A}, {B}$) Simulated speed-of-sound distributions $c(\mathbf{r})$. ({C}, {D}) Optimized wave velocity $\hat{c}(\mathbf{r})$. (${E}, {F}$) Estimation of the local speed-of-sound map ${c}(\mathbf{r})$ {with each pixel reassigned to its estimated position.} (${G}, {H}$) Original ultrasound image. (${I},{J}$) Corrected image with each pixel reassigned to its estimated position.}
    \label{figsupp_simu}
\end{figure*}

To extend and validate our approach for a heterogeneous sound speed distribution, \corr{new ultrasound data sets} have been computed with k-Wave~\cite{treeby_k-wave_2010}  {(Appendix~\ref{J})}, a time domain simulation software based on the k-space pseudo-spectral method. The considered speed-of-sound distributions are layered media with parallel (Fig.~\ref{figsupp_simu}A) and oblique interfaces (Fig.~\ref{figsupp_simu}B) with respect to the ultrasonic probe. Short-scale fluctuations of density have been superimposed to generate a random speckle characteristic of ultrasound imaging in soft tissues. A set of point-like targets is also included to quantify the spatial resolution at different locations.

Ultrasound images (Eq.~\ref{confocal}) are computed from the corresponding reflection matrices by considering a homogeneous wave velocity model $c_0=1540$ m.s$^{-1}$ corresponding to the speed-of-sound {in the second layer}. The result is displayed in Figs.~\ref{figsupp_simu}G and H. As expected, the mismatch between $c(x,z)$ and $c_0$ results in a shift of the interface and target locations with respected to their true position. The images of the bright targets also shows the impact of an inexact wave velocity model on the spatial resolution, with bright point-like scatterers appearing as curved arches. 

On the contrary, as already shown with the phantom experiment, the wrong velocity model has not a clear impact on the speckle grain size. Nevertheless, a coherent RPSF can be determined for each speckle grain $\mathbf{r}_{\textrm{in}}=(x_{\textrm{in}},t)$ by computing the SVD of the de-scanned matrix (Eqs.~\ref{SVD} and \ref{PSFcoh}) over sliding spatial windows {$\mathcal{P}$}  {(Appendix~\ref{N})}. An optimized wave velocity $\hat{c}(x,t)$ is determined for each spatio-temporal point $(x,t)$ by maximizing the corresponding RPSF. The resulting maps $\hat{c}(x,z_t)$ are displayed in Figs.~\ref{figsupp_simu}C and D for each configuration. They are far from the ground truth distributions displayed in Figs.~\ref{figsupp_simu}A and B. Indeed, the optimized wave velocity is not an estimator for the local speed-of-sound but for the inverse of the mean slowness ${\bar{s}(x,t)}$, averaged between the probe surface and the focusing point $(x,t)$: 
\begin{equation}
\label{integrate}
{\bar{s}(x,t)}=\frac{1}{z_t(x)}\int_0^{z_t(x)} \frac{dz}{c(x,z)},
\end{equation}
with $z_t(x)=\hat{c}(x,t)t/2$, the depth of the isochronous volume for each echo time $t$ and lateral position $x$. Equation~\ref{integrate} is extremely simplified since it only takes into account vertical paths, thereby neglecting refraction phenomena. Nevertheless, a numerical inversion of Eq.~\ref{integrate} can be performed to retrieve an estimator of the local speed-of-sound $c(x,z)$ from $\hat{c}(x,t)$~\cite{jakovljevic_local_2018} (Appendix~\ref{N}). The resulting speed-of-sound maps are displayed in Figs.~\ref{figsupp_simu}E and F. 

\corr{For a flat interface (Figs.~\ref{figsupp_simu}E), a close agreement is found with the ground-truth (Fig.~\ref{figsupp_simu}A). The lateral invariance of this first configuration makes the assumption leading to Eq.~\ref{integrate} actually valid.} The estimation of $c(x,z)$ \corr{is therefore} reliable, with a mean error $\delta c$ in the tissue layer of the order of 10 m.s$^{-1}$. The axial resolution $\delta z$ can be estimated by investigating the axial dependence of ${c}(x,z)$. In the present case, we find {$\delta z \sim 5$ mm}. 

\corr{Not surprisingly, the numerical inversion of Eq.~\ref{integrate} is less robust for an oblique interface between the muscle and tissues behind. Important fluctuations of the speed-of-sound are actually observed after the interface (Fig.~\ref{figsupp_simu}F). 
Thus, our approach cannot provide a quantitative map of the local speed-of-sound for complex speed-of-sound distributions. However,} it allows a direct compensation of axial aberrations in the ultrasound images. Indeed, the isochronous volume and each focal plane can be matched by reassigning to each point $(x,z)$ of the medium a correct echo time based on our estimation of the depth-averaged slowness, such that
\begin{equation}
\label{confocal_corr}
\mathcal{I}'(x,z)=R({x,x},t=2z/{\hat{c}(x,z),\hat{c}(x,z)}).
\end{equation}
Contrary to the original ultrasound image whose axial dimension was dictated by the echo time and wave velocity assumption $c_0$, this new ultrasound image displays the medium as a function of the real depth $z$. Moreover, the use of the depth-averaged velocity enables the compensation of defocus. The resulting images are shown in Figs.~\ref{figsupp_simu}I and J for the numerical experiments described above. Compared to their original versions (Figs.~\ref{figsupp_simu}G and H), those two images show several striking improvements: (\textit{i}) A drastic contrast enhancement \corr{by $10$ dB at shallow depth}; \corr{(\textit{ii}) A fine compensation of defocus highlighted by an interface between the two layers that now emerges at the correct depth;}
(\textit{iii}) A drastic gain in transverse resolution highlighted by the image of the bright targets. After this numerical validation, the \corr{application} of the method \corr{to in-vivo liver is investigated.} 

\subsection{\corr{Application to} a pathological clinical case}

\corr{To illustrate the potential of our method for medical diagnosis,} a pathological liver case is now addressed in vivo. More precisely, we target a patient liver, which is difficult to image due to an irregular arrangement of adipose and muscle tissues upstream of the liver. Moreover, this patient is potentially suffering from steatosis.
This disease corresponds to an accumulation of fat droplets in the liver that induces a low speed-of-sound and enhanced scattering. 
While this disease can manifest as a bright speckle~\cite{mehta_non-invasive_2008,dasarathy_validity_2009}, this observable is only qualitative and operator-dependent. 
The effectiveness of ultrasound for diagnosing hepatic steatosis is reduced in obese patients~\cite{Almeida2008}. Indeed, because the ultrasonic waves must travel through successive layers of skin, fat, and muscle tissue before reaching the liver, both the incident and reflected wave-fronts undergo strong aberrations~\citep{Hinkelman1998,Browne2005} and multiple scattering (clutter noise)~\citep{Lediju2009}. 
Hence, there is a strong need to overcome aberrations
for the early detection of such a disease.
\begin{figure*}
\centering
\includegraphics[width=1\linewidth]{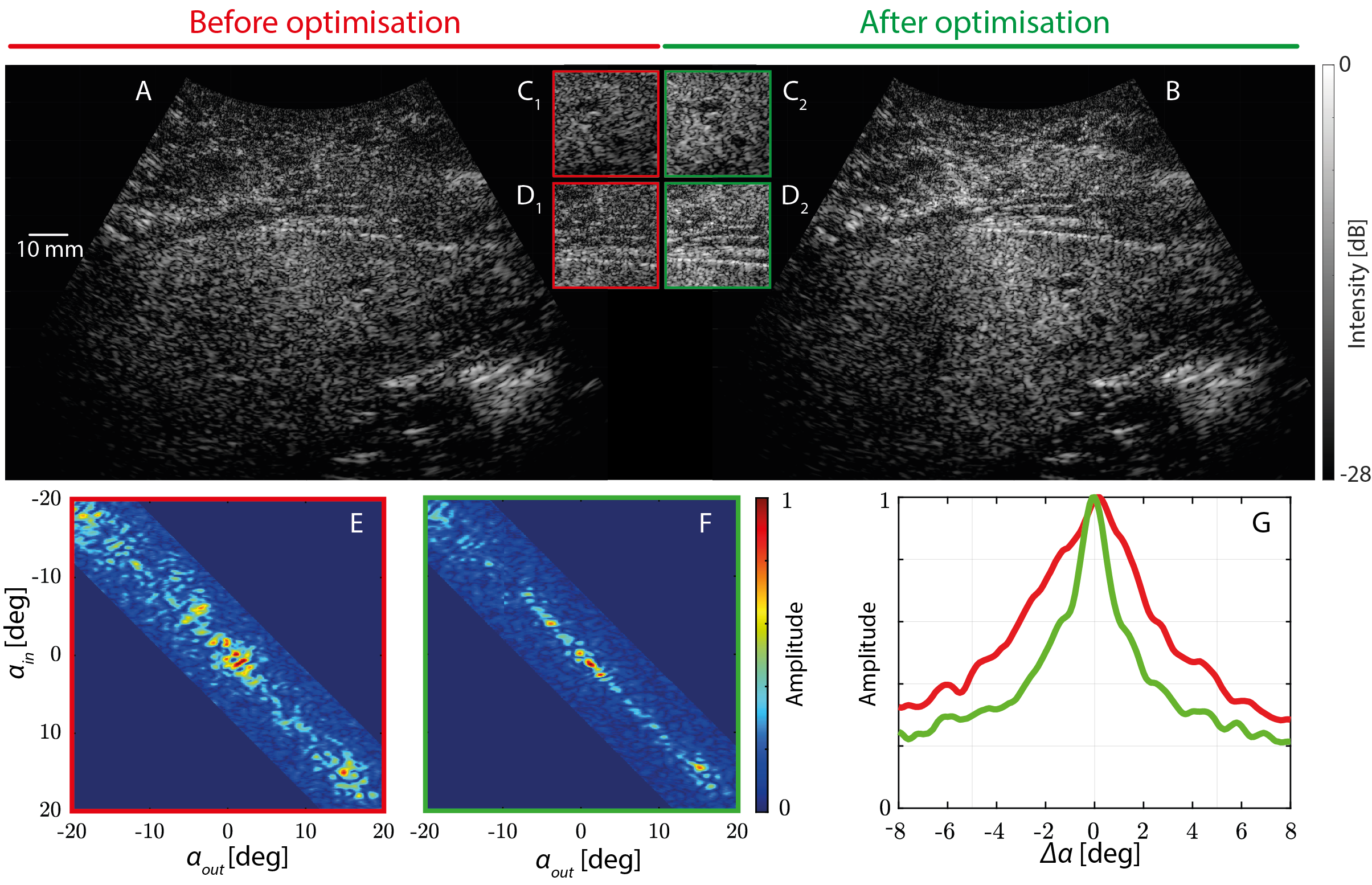}
\caption{\textbf{Speed of sound optimization in the liver experiment.} {(A, B)} Original and optimized ultrasound images, respectively. Both images are normalized by the global maximum between the two images and are displayed along the same depth axis, which is estimated with a constant speed of sound ($c_0=1540$ m/s). (B,C) Zoom on specific areas of the field-of-view containing either muscle fibers or veins before and after optimization, respectively. Subscripts ``1'' and ``2'' refer to two different areas of the field of view. (E,F) Focused reflection matrix corresponding to $t=90.9$ $\mu$s {($\rho_0=c_0t/2 \sim70$mm)} before and after optimization respectively. (G) Incoherent RPSF before (red curve) and after correction (green curve). }\label{fig_invivo_OptimisationImage}
\end{figure*}

\corr{This in vivo liver ultrasound dataset is extracted from an observational and retrospective, bicentric study (Perpignan Hospital and Angers University Hospital) performed in conformation with the declaration of Helsinki and that was approved by an ethics committee (EC). The reflection matrix has been recorded with a curved array of transducers {(XC 6-1, Supersonic Imagine)} whose characteristics are provided in Tab.~\ref{AcquisitionParam}. A set of diverging waves is generated by applying the same time delay that we would apply to generate a plane wave from a linear array. The beamforming algorithm is equivalent to Eq.~\ref{confocal} except that Cartesian coordinates $(x,z)$ are replaced by polar ones $(\alpha,\rho)$ (Appendix~\ref{L}).}

The conventional image (Eq.~\ref{confocal}, $c_0=1540$ m/s) is displayed in Fig. \ref{fig_invivo_OptimisationImage}A. It shows a poor contrast due to the aberrations induced by the adipose and muscle tissues at shallow depths. This poor image quality is confirmed by investigating the focused reflection matrix at a given echo time $t=91$ $\mu$s (Fig.~\ref{fig_invivo_OptimisationImage}E). While the focused reflection matrix ${\mathbf{R}_{{\alpha\alpha}}}(z)$ shall be nearly diagonal in an ideal case~\cite{lambert_reflection_2020}, it here displays a spreading of the back-scattered energy over off-diagonal coefficients. This feature is a manifestation of: (\textit{i})  aberrations induced by the mismatch between $c_0$ and $c(\mathbf{r})$; (\textit{ii}) multiple scattering events taking place upstream of the focal plane~\cite{lambert_reflection_2020,lambert_ultrasound_2022}. This observation is confirmed by investigating the transverse dependence of the incoherent PSF shown in  Fig. \ref{fig_invivo_OptimisationImage}G (red curve). This RPSF displays the following shape: A single scattering peak enlarged by aberrations on top of a multiple scattering background whose weight is far from being negligible, since it reaches the value of 20\%.

\begin{figure*}
\includegraphics[width=1\linewidth]{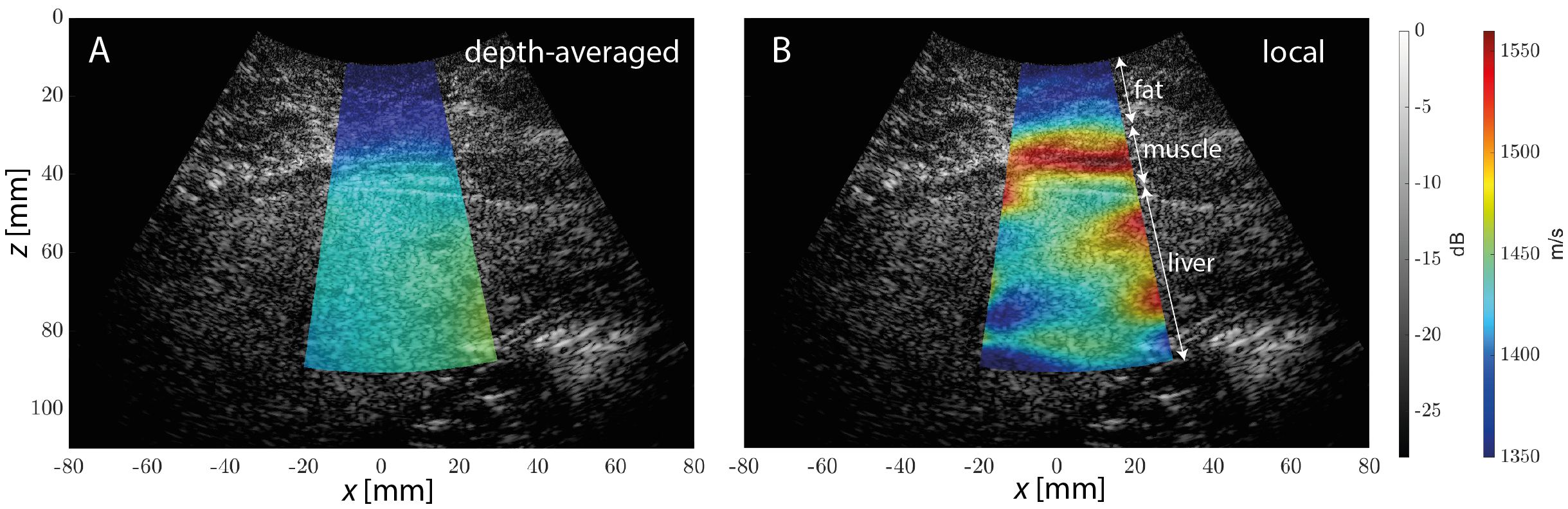}
    \caption{\textbf{Speed-of-sound mapping in the liver experiment.} (A) Depth-averaged and (B) local speed-of-sound reconstruction superimposed onto the standard image of the liver reflectivity. Both images are displayed with an estimated depth axis corresponding to a constant speed-of-sound, such that $\rho_0=c_0t/2$ with $c_0=1540$ m/s.}
    \label{fig_invivo_SOS}
\end{figure*}
Following the method described above, this observable can be exploited for speed-of-sound tomography by scanning $c_0$. An optimized wave velocity $\hat{c}(x,z_t)$ can be retrieved for each pixel of the ultrasound image, and the resulting map is shown in Fig.~\ref{fig_invivo_SOS}A. The optimal speed-of-sound $\hat{c}(x,z_t)$ start from a very low value ($\sim1400$ m/s) at shallow depth before suddenly increasing at $z_t$=35 mm and reaching a plateau ($\sim1480$ m/s) beyond $z_t=40$ mm. If such a map can be useful to improve the ultrasound image, as we will see further, \corr{it is not representative of} the speed-of-sound distribution since the probed velocity is averaged from the probe to each focusing point. As before, an inversion of Eq.~\ref{integrate} is needed (Appendix~\ref{N}) to provide \corr{a qualitative map} of the local speed-of sound. The result is displayed in Fig \ref{fig_invivo_SOS}B. It clearly highlights the presence of three tissue layers: (\textit{i}) adipose tissue, from $z=0$ to 30 mm, with a low speed-of-sound $c\sim1400$ m/s; (\textit{ii}) muscle tissue, from $z=30$ to $z=40$ mm, which induce a sudden increase in the speed-of-sound ($c \sim 1550$ m/s); (\textit{iii}) the liver, beyond $z=40$ mm, which is characterized by \corr{an extremely} slow speed-of-sound ({$c \sim 1480$ m/s in average}). The speed-of-sound map is coherent with the different features shown by the ultrasound image, with a heterogeneous speckle in the fat layer, muscle fibers in the intermediate region and a homogeneous speckle in the liver. However, this map not only confirms the structural information provided by the ultrasound image, it also provides a \corr{qualitative} measurement of $c$ \corr{consistent with a proton density fat fraction rate (PDFF) of 20\% in liver~\cite{imbault_robust_2017}}. Our measurement thus indicates that the patient is likely to suffer from steatosis.

Beyond this crucial information, mapping the speed-of-sound leads to a more contrasted ultrasound image $\mathcal{I'}$(Fig.~\ref{fig_invivo_OptimisationImage}B), as shown by the large improvement of the speckle brightness compared to its original version (Fig.~\ref{fig_invivo_OptimisationImage}A).  The interfaces between tissues show a much better lateral coherence [see comparison between insets shown in Figs.\ref{fig_invivo_OptimisationImage}C and D]. This is especially the case at shallow depths, where the variations of the sound velocity are the most drastic and their impact on the image the most important. The correction of {axial aberrations} is also accompanied by a drastic reduction in transverse aberrations, and thus a significant improvement in terms of resolution. Thus, it allows better visualization of structures such as muscle fibers or veins inside the liver. Such resolution enhancement can be quantified by considering the distribution of energy inside the focused reflection matrix. After defocus compensation, most of the back-scattered energy is brought back in the vicinity of the diagonal coefficients [see comparison between Figs.~\ref{fig_invivo_OptimisationImage}E and F]. A resolution enhancement of about a factor of two is highlighted by comparing the transverse spreading of the confocal peak exhibited by the incoherent {RPSF} before and after defocus correction [Fig.~\ref{fig_invivo_OptimisationImage}G]. The contrast enhancement is also shown by the lower multiple scattering background observed after correction. 

\begin{figure}[t!]
\includegraphics[width=12cm]{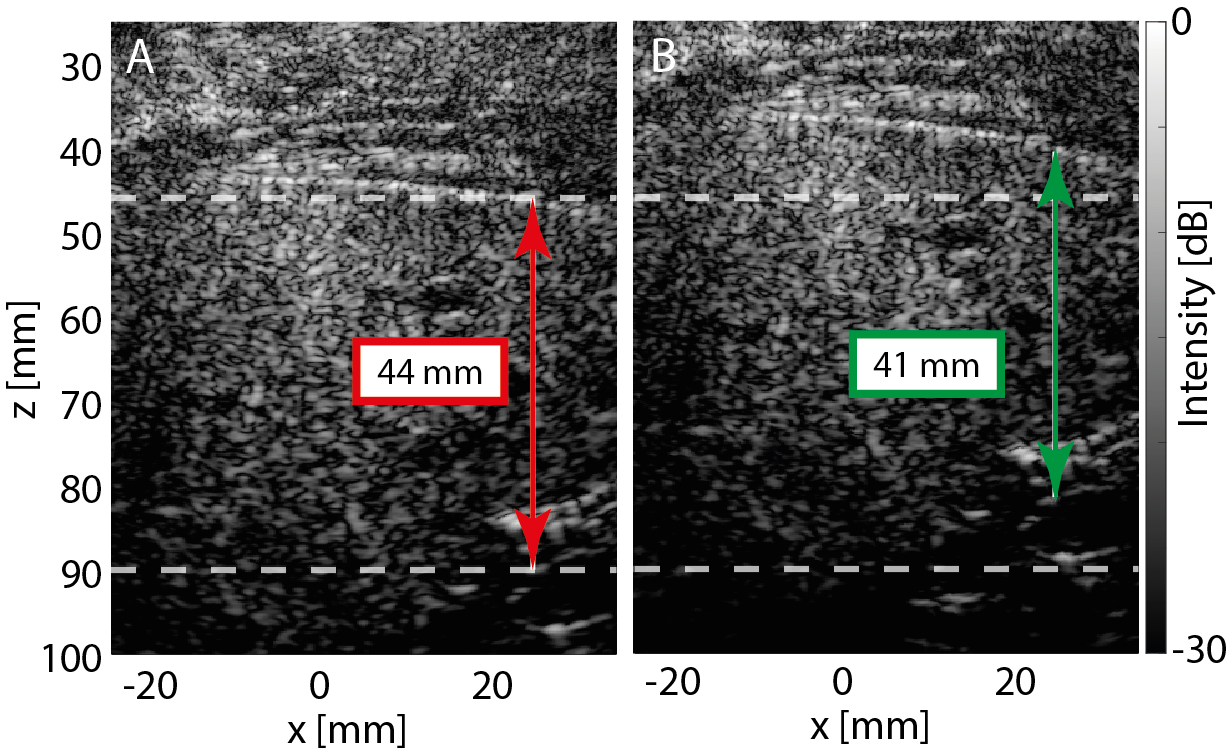}
    \caption{\textbf{Depth axis rescaling of the liver image.} {(A)} Ultrasound image displayed with a depth axis estimated with a constant speed of sound such that ${\rho_0=c_0t/2}$ with $c_0=1540$ m/s. {(B)} Final image (Eq.~\ref{confocal_corr}) displayed with a depth axis rescaled by the depth-averaged wave velocity such that the depth $\rho(x,t)$ of each initial pixel is re-assigned to $\hat{c}(x,t)t/2$.}
    \label{fig_distances}
\end{figure}

Compared with the ultrasound image (Fig.~\ref{fig_invivo_OptimisationImage}A) whose axial dimension is dictated by the echoes' time-of-flight, each pixel in the optimized image is shifted to its real position in depth, thereby giving access to absolute distances. This feature can be a major breakthrough in ultrasound imaging since a lot of diagnoses rely on distance measurements~\citep{Scorza2015} as, for instance, in obstetrics to monitor fetal growth or detect chromosomal abnormality~\citep{Nicolaides1994,Snijders1998}. As an example, we consider the distance between two speckle spots at the extremity of the red and green arrows in Fig.~\ref{fig_distances}. The distance between those two points is overestimated by 3 mm in the initial image (Fig. \ref{fig_distances}A) compared with the optimized image (Fig. \ref{fig_distances}B). This difference stems from the re-scaling of the depth axis operated under our approach. This observation highlights the benefit that could provide a depth reassignment of pixels for ultrasound diagnosis.

\section{Discussion}

In this experimental proof-of-concept, we demonstrated the capacity of UMI to \corr{exploit speckle for a local optimization of the speed-of-sound and a compensation of axial aberrations in reflection}. This work is not only an extension of previous studies, since several crucial elements have been introduced to make our approach more robust.

Compared with previous works that relied on a maximization of the image quality~\cite{benjamin_novel_2018,zubajlo_experimental_2018,napolitano_sound_2006,cho_efficient_2009,yoon_optimal_2012,yoon_vitro_2011,shin_estimation_2010,perrot_so_2021}, or parameters such as coherence~\cite{anderson_direct_1998,imbault_robust_2017,ali_local_2021} or focusing factor~\cite{lambert_reflection_2020,lambert_ultrasound_2022a}, our approach is based on a coherent average {(SVD)} of the focusing process over different speckle grains. Such a process is more robust with \corr{respect to} multiple scattering and noise that tend to vanish with coherent averaging.  Moreover, the access to the Gouy phase allows us to reduce the bias of the speed-of-sound estimator by a factor of about three compared to usual observables relying on an intensity maximization. \corr{Note also that a recent theoretical study~\cite{Garnier2025} provided a rigorous mathematical proof of our approach. It also generalizes our theoretical predictions that have been derived in this work under a Gaussian beam approximation.}

While our method is extremely relevant for axial aberration correction, it remains perfectible for mapping the transverse variations of the speed-of-sound. Indeed, Eq.~\ref{integrate} only takes into account vertical paths and neglects refraction phenomena. Nevertheless, our approach can be complementary \corr{with other methods such as CUTE}~\cite{jaeger_computed_2015,stahli_improved_2020,Staehli2023} and related approaches~\cite{heriard-dubreuil_refraction-based_2023}. A problem in CUTE is the residual bias due to the initial speed-of-sound hypothesis. Here, by scanning the model wave velocity $c_0$, our approach provides an unbiased estimation of the depth-averaged speed-of-sound. \corr{This information could therefore be used by CUTE or equivalent approaches to reduce the bias with respect to the initial speed-of-sound model.} 

\corr{The inversion scheme used to derive the local sound velocity map remains largely perfectible. Further efforts are needed to define a more complex inverse problem taking into account oblique paths and their distorted trajectories due to refraction~\cite{beuret_refraction-aware_2020,heriard-dubreuil_refraction-based_2023}.} 
In that respect, the inversion problem can be solved iteratively by updating the forward model. As a first step, the estimated sound speed map can actually be used to build a more complex beamforming scheme accounting for refraction phenomena~\cite{jaeger_full_2015,ali_distributed_2022}. This process can then be iterated using a differential beamformer~\cite{Simson2023} that optimizes the sound speed distribution using a variety of physical constraints based on speckle brightness, coherence maximization or RPSF optimization as in the current paper. Such an iterative process can lead to a sharper estimation of the speed-of-sound and a close-to-ideal ultrasound image, {not only} in terms of transverse and axial resolution {but also by {a} correct positioning of scatterers {in depth}}.

\corr{Note also that the method presented in this paper is very flexible and can actually include more sophisticated wave propagation models. Depending on the experimental configuration we have to cope with, one can actually choose other parameters than the speed-of-sound to optimize the focusing laws. In a multi-layered medium, an optimization of the focusing depth can directly compensate for the defocus induced by upstream layers of different wave velocity~\cite{Bureau}. This approach can be particularly fruitful for compensating the strong axial aberrations induce by the skull in trans-cranial imaging. However, whether it be for liver or brain imaging, reverberation phenomena can generate strong artifacts on the ultrasound image and hamper the measurement of the speed-of-sound. In the third paper of the series, we will show how we can extract the movie of the reverberated wave-field by tuning the echo time rather than the wave speed. Based on this observable, we will be able to tailor spatio-temporal focusing laws in order to harness the multiply-reflected paths and make them interfere constructively with the direct path at any focusing point.}

\section{Conclusion}

\corr{The self-portrait of wave focusing in speckle provided by matrix imaging is very flexible. In this paper, we showed how it can enable an axial compensation of aberrations. This is done by exploiting the Gouy phase shift exhibited by the focused wave-field when the wave velocity model is optimal. Besides directly generating images with better contrast and higher resolution, it allows positioning the depth of each scatterer with greater accuracy and thus to better evaluate the distance across ultrasound images. The differentiation of the obtained depth-averaged slowness can provide a qualitative representation of the local speed-of-sound distribution. We have demonstrated the potential benefit of our approach in a pathological clinical case, a liver of a difficult-to-image patient suffering from steatosis. Coupled with more quantitative imaging methods~\cite{Staehli2023,Simson2023} that are more sensitive to initial conditions,  our approach could be rewarding for quantification of the sound velocity, a bio-marker not only for liver disease but also for tumor assessment \cite{Gordon1995,ozmen_comparing_2015,ruby_breast_2019}. Moreover, matrix imaging can apply to any kind of waves for which the multi-element technology exists~\cite{balondrade_multi-spectral_2023,giraudat_unveiling_2023}. Mapping the 3D distribution of the optical index in tissues~\cite{chen_multi-layer_2020} or elastic wave speeds in non-destructive testing~\cite{Bazulin2022} and reflection seismology~\cite{Shiraishi2023} are all examples of relevant applications for the general method proposed in this paper.}




\noindent\textbf{Acknowledgments.}
The authors are grateful for the funding provided by the European Research Council (ERC) under the European Union's Horizon 2020 research and innovation program (grant agreement 819261, REMINISCENCE project, AA). E.B. and W.L. acknowledges financial support from the SuperSonic Imagine company. N.K.M. acknowledges funding by LABEX WIFI (Laboratory of Excellence within the
French Program Investments for the Future; ANR-10-LABX-24 and ANR-10-
IDEX-0001-02 PSL*, M.F.). \\

\noindent\textbf{Author contributions.}
A.A. and M.F. initiated the project. A.A. supervised the project. F.B. performed the phantom experiment. F.B. performed the numerical simulations. L.C. and A.G. performed the liver experiment. F.B., E.B., N.K.M. and W.L. developed the post-processing tools. A.A. performed the theoretical analysis.  F.B. and A.A. prepared the manuscript. F.B., E.B., N.K.M., E.G., A.L.B., W.L., L.C., A.G., M.F. and A.A. discussed the results and contributed to finalizing the manuscript. 

\clearpage

\appendix
\section{Delay-and-sum algorithm - linear array}
\label{K}
For a linear array, the general procedure to build the confocal image and the focused reflection matrix is the delay-and-sum (DAS) algorithm described in Eqs.~\ref{confocal} and \ref{focusedR}. The time-of-flight $\tau_{\textrm{in}}$ describes the travel time for each incident plane wave $\theta_\textrm{in}$ from the probe to any targeted focusing point $({x}_\textrm{in},z_0=c_0t/2)$:
\begin{equation}
\tau_{\textrm{in}}(\theta_\inm,\corr{x_\inm},t,c_0)=\frac{\corr{x_\inm}\sin({\theta^0_\inm})+(c_0t/2)\cos({\theta^0_\inm})}{c_0} + \tau_0(\theta_\inm,c_\textrm{acq}).
\end{equation}
with ${\theta^0_\inm}$ the incident angle of the emitted plane wave $\theta_\textrm{in}$ for a wave velocity model $c_0$, such that:
\begin{equation}
    {\theta^0_\inm}=\arcsin\left(\frac{c_0}{c_\textrm{acq}}\sin(\theta_\inm)\right).
\end{equation}
The additional time delay $\tau_0(\theta_\inm)$ corresponds to the time shift applied to each incident plane wave in order to set the same time origin for each insonification~\cite{montaldo_coherent_2009}. This time origin corresponds to the time when the incident pulse is emitted by the central element of the array.

The time-of-flight $\tau_{\textrm{out}}$ in Eqs.~\ref{focusedR} and \ref{confocal} is the travel time for the reflected wave from the focusing point $({x}_\textrm{out},z_0=c_0t/2)$ to any transducer $u_\outm$ of the probe: 
\begin{equation}
\tau_\outm (u_\outm,x_{\textrm{out}},t,c_0)=\frac{\sqrt{\left(x_\outm-u_\outm\right)^2+\left(c_0t/2\right)^2}}{c_0}.
\end{equation}

\section{Theoretical expression of the focused reflection matrix}
\label{B}
The reflection matrix $\mathbf{R}_{uu}(t)$ expressed in the transducer basis can be decomposed in the temporal Fourier domain as follows:
\begin{equation}
\mathbf{R}_{uu}(t)=\int d\omega \overline{\mathbf{R}}_{uu}(\omega) \exp(i\omega t)
\end{equation}
Under a single scattering assumption, the coefficients of the monochromatic reflection matrix $\overline{\mathbf{R}}_{{uu}}(\omega)$ can be expressed as follows:
\begin{equation}
\label{Ruu}
\overline{R}({u}_\textrm{out},{u}_\textrm{in},\omega)= \int d\mathbf{r}G_0(\mathbf{u}_\textrm{out},\mathbf{r},\omega)\gamma(\mathbf{r}) G_0(\mathbf{u}_\textrm{in},\mathbf{r},\omega)
\end{equation}
with $\gamma(\mathbf{r})$, the medium reflectivity and $G_0(\mathbf{u},\mathbf{r})$, the {free space} Green's function that accounts for propagation of a monochromatic wave between the transducer located at $\mathbf{u}=(u,0)$ and any point $\mathbf{r}=(x,z)$ inside the medium. In a 2D configuration, the wave equation Green's function reads
\begin{equation}
 G_0 ( \mathbf{u},\mathbf{r} ,\omega) = -\frac{i}{4} \mathcal{H}_0^{(1)} \left (k \sqrt{(x-u)^2+z^2} \right )
\label{eq:chap2_solutionGreen}
\end{equation}
with $\mathcal{H}_0^{(1)}$ the Hankel function of the first kind and $k=\omega/c$, the wave number. In the far-field, the asymptotic expression of the Green's function is:
\begin{equation}
	G_0( \mathbf{u} , \mathbf{r} ,\omega) \approx \frac{-e^{i \pi / 4}}{\sqrt{8 \pi k} \left [{(x-u)^2+z^2} \right ]^{1/4} }  \exp \left( i {k}\sqrt{(x-u)^2+z^2}  \right).
	\label{eq:G0champlointain}
\end{equation}

A spatial Fourier transform at input and output of $\mathbf{R}_{uu}(\omega)$ leads to a reflection matrix $\overline{\mathbf{R}}_{\kappa \kappa}(\omega,z=0)$ expressed in the plane wave basis at the medium surface ($z=0$):
\begin{equation}
\overline{\mathbf{R}}_{{\kappa \kappa}}(\omega,z=0)=\mathbf{F}_{\kappa u} \times  \overline{\mathbf{R}}_{uu}(\omega) \times \mathbf{F}_{\kappa u}^\top 
\end{equation}
with $\mathbf{F}=[F({\kappa},{u})]$, the Fourier transform operator, such that
\begin{equation}
F({\kappa},u)=\exp \left ( -i {\kappa} u\right )
\end{equation}
In the plane wave basis, Eq.~\ref{Ruu} becomes:
\begin{equation}
\overline{R}({\kappa}_{\textrm{out}},{\kappa}_{\textrm{in}},\omega,z=0)=\int dz \,P({\kappa}_{\textrm{out}},z) P({\kappa}_{\textrm{in}},z) \Gamma({\kappa}_{\textrm{in}}+{\kappa}_{\textrm{out}},z)
\end{equation}
with {$\Gamma({\kappa}_{\textrm{in}}+{\kappa}_{\textrm{out}},z)= \int dx \gamma(x,z) \exp \left [ -i ({\kappa}_{\textrm{out}}+{\kappa}_{\textrm{in}})x \right]$}, the Fourier transform of the object reflectivity. $\mathbf{P}_{\kappa}(\omega,z,c)=[P({\kappa},\omega,z,c)]$ is the propagator describing plane wave propagation between the probe ($z=0$) and the plane $z$ inside the medium, such that
\begin{equation}
P({\kappa},z,c)= 
         \exp(iz\sqrt{k^2-|{\kappa}|^2})
\label{eq:chap2_solutionGreen2}
\end{equation}
for $|\kappa|<k$ and zero elsewhere if $z>>\lambda$.

The focused reflection matrix $\mathbf{R}_{xx}(t,c_0)$ defined in Eq.~2 can be decomposed in the temporal Fourier domain as follows:
\begin{equation}
\label{TFt}
\mathbf{R}_{xx}(t,c_0)= \int_{\omega_-}^{\omega_+} d\omega \overline{\mathbf{R}}_{xx}(\omega,z_0,c_0) \exp(i\omega t)
\end{equation}
with $\omega_{\pm}=\omega_c \pm \delta \omega/2$, $\omega_c$, the central frequency and $\delta \omega$, the frequency bandwidth. Each monochromatic focused reflection matrix $ \overline{\mathbf{R}}_{xx}(\omega,z_0,c_0)$ results from a focusing process at the expected ballistic depth $z_0=c_0t/2$~\cite{lambert_reflection_2020}. $ \overline{\mathbf{R}}_{xx}(\omega,z_0,c_0)$ can be projected in the plane wave basis as follows:
\begin{equation}
\label{TFs}
\overline{\mathbf{R}}_{\kappa \kappa}({\omega,z_0},c_0) = \mathbf{F}_{kx} \times    \overline{\mathbf{R}}_{xx}(\omega,z_0,c_0) \times \mathbf{F}_{kx}^{\top}.
\end{equation}
Under a matrix formalim, the delay-and-sum beamforming process (Eq.~2) can be expressed in the Fourier domain as the following Hadamard product:
\begin{equation}
\overline{ \mathbf{R}}_{{\kappa \kappa}}(\omega,{z_0},c_0)= \exp \left ( 2 i \omega  z_0 /c_0\right ) \mathbf{P}^*(z_0,\omega,c_0) \circ \overline{\mathbf{R}}_{{\kappa \kappa}}(z=0,\omega) \circ \mathbf{P}^\dag(z_0,\omega,c_0),
\end{equation}
which can be expressed in terms of matrix coefficients as follows:
\begin{align}
 \overline{R}({\kappa}_{\textrm{out}},{\kappa}_{\textrm{in}},z_0,\omega,c_0) = & \exp \left ( 2 i \omega  z_0 /c_0\right )  \\\
 & \times \int dz   \Gamma(\bm{\kappa}_{\textrm{in}}+\bm{\kappa}_{\textrm{out}},z) O_{\textrm{out}}(\bm{\kappa}_{\textrm{out}},z) O_{\textrm{in}}(\bm{\kappa}_{\textrm{in}},z)\\
 & \times \exp \left (- i z_0 \sqrt{\left ( \frac{\omega}{c_0} \right)^2-|\bm{\kappa}_{\textrm{out}}|^2} \right ) \exp \left (+ i z \sqrt{\left ( \frac{\omega}{c} \right)^2-|\bm{\kappa}_{\textrm{out}}|^2} \right ) \\
 & \times  \exp \left (- i z_0 \sqrt{\left ( \frac{\omega}{c_0} \right)^2-|\bm{\kappa}_{\textrm{in}}|^2} \right ) \exp \left (+ i z \sqrt{\left ( \frac{\omega}{c} \right)^2-|\bm{\kappa}_{\textrm{\corr{in}}}|^2} \right ),
\end{align}
where the functions $O_{\textrm{in}}$ and $O_{\textrm{out}}$ account for the angular aperture applied at input and output during the beamforming process, respectively. Under the paraxial approximation, this last equation can be rewritten as follows:
\begin{align}
 \overline{R}({\kappa}_{\textrm{out}},{\kappa}_{\textrm{in}},z_0,\omega,c_0) = &  \int dz   \exp \left ( 2 i \omega  z /c\right ) \Gamma({\kappa}_{\textrm{in}}+{\kappa}_{\textrm{out}},z) O_\textrm{out}({\kappa}_{\textrm{out}},z) O_\textrm{in}({\kappa}_{\textrm{in}},z)\\
 & \times \exp \left (- i  \frac{|{\kappa}_{\textrm{out}}|^2+|{\kappa}_{\textrm{in}}|^2}{2\omega} (c z- c_0 z_0) \right ) .
\end{align}
The last expression can be recast as follows
\begin{equation}
\label{Rkk}
 \overline{R}({\kappa}_{\textrm{out}},{\kappa}_{\textrm{in}},z_0,\omega,c_0) =   \int dz   \exp \left ( 2 i \omega  z /c\right ) \Gamma({\kappa}_{\textrm{in}}+{\kappa}_{\textrm{out}},z) H_\textrm{out}({\kappa}_{\textrm{out}},z) H_\textrm{in}({\kappa}_{\textrm{in}},z).
\end{equation}
with $H_\textrm{in}$ and $H_\textrm{out}$, the input and output transfer functions of the imaging process:
\begin{equation}
H_\textrm{in/out}({\kappa},z,z_0,c_0) =  O_\textrm{in/out}({\kappa},z)\exp \left (- i  \frac{|{\kappa}|^2}{2\omega} (c z- c_0 {z_0}) \right ).
\end{equation}
The cancellation of the phase term defines the position $z_f$ of the focusing plane, such that
\begin{equation}
z_f=\frac{c_0}{c}z_0. 
\end{equation}
In first approximation, one can consider the transfer function as relatively constant over the frequency bandwidth and consider its value at the central frequency $\omega_c$, such that: 
\begin{equation} 
\label{approxH}
H_\textrm{in/out}({\kappa},z,z_0,c_0) \simeq O_\textrm{in/out}({\kappa},z)\exp \left (- i  \frac{|{\kappa}|^2}{2\omega_c} (c z- c_0 {z_0}) \right )
\end{equation}
Injecting Eq.~\ref{Rkk} into Eqs.~\ref{TFs} and \ref{TFt} leads to the following expression for the focused reflection matrix coefficients:
\begin{equation}
\label{R3}
R(x_\textrm{out},x_\textrm{in},t,c_0)= \int d\omega  \int dz \int d\corr{x} \exp \left [ i{\omega (t-2z/c)} \right ] h_\textrm{out}(x-x_\textrm{out},z,z_0,c,c_0) \gamma(\corr{z},z) h_\textrm{in}(x-x_\textrm{in},z,z_0,c,c_0) 
\end{equation}
where $h_\textrm{in/out}$ is the Fourier transform of the transfer function $H_\textrm{in/out}$: 
\begin{equation}
\label{impulse}
h_\textrm{in/out}(x,z,z_0,c_0)=\int d {\kappa} H_\textrm{in/out}({\kappa},z,z_0,c_0) \exp(-i {\kappa}x).
\end{equation}
$h_\textrm{in/out}(x,z,z_0,c_0)$ corresponds to the point-spread function of the imaging system at depth $z$ when trying to focus at plane $z_0$ assuming a wave velocity model $c_0$. 

Due to the broad spectrum of ultrasound signals, the integral over frequency in Eq.~\ref{R3} can be simplified in first approximation as follows:
\begin{equation}
R(x_\textrm{out},x_\textrm{in},t,c_0)  \simeq  \int dz \delta(t-2z/c) \int d{x} h_\textrm{out}(x-x_\textrm{out},z,z_0,c,c_0) \gamma(x,z) h_\textrm{in}(x-x_\textrm{in},z,z_0,c,c_0).
\end{equation}
The Dirac distribution $\delta$ in the last expression accounts for the time gating operation and implies that, for a given time-of-flight $t$, the contribution of the scattered wave-field is induced by the set of scatterers lying in the vicinity of the isochronous plane located at $z_t=ct/2$. The expression of the focused $\mathbf{R}-$matrix coefficients becomes
\begin{equation}
\label{focR}
R(x_\textrm{out},x_\textrm{in},t,c_0) \corr{\simeq} \int dx \, h_\textrm{out}(x-x_\textrm{out},z_t,z_0,c_0) \gamma(x,z_t) h_\textrm{in}(x-x_\textrm{in},z_t,z_0,c_0),
\end{equation}
or, expressed in the de-scanned basis,
\begin{equation}
\label{eqR0}
R_D(\lbrace\Delta x,c_0 \rbrace,\lbrace x_\textrm{in},t \rbrace)  \propto  \int dx' \, h_{\textrm{out}}(x'-\Delta x,z_t,z_0,c_0) \gamma(x'+x_{\textrm{in}},z_t) h_{\textrm{in}}(x',z_t,z_0,c_0).
\end{equation}

\section{Incoherent RPSF}
\label{D}
In the accompanying paper, the focusing quality is first assessed by considering an incoherent average of each column of $\mathbf{R}_D$. Using Eq.~\ref{eqR0}, the incoherent intensity, $RPSF_\textrm{inc}^2(\Delta x,c_0)=\left \langle |R_D(\lbrace\Delta x,c_0 \rbrace,\lbrace x_\textrm{in},t \rbrace) |^2 \right \rangle$, can be expressed as follows:
\begin{align}
RPSF_{\textrm{inc}}^2(\Delta x,c_0) \propto   \Biggl  \langle \int dx' \int dx'' \, & h_{\textrm{out}}(x'-\Delta x,z_t,z_0,c_0) h^*_{\textrm{out}}(x''-\Delta x,z_t,z_0,c_0)   \nonumber \\
& \gamma(x'+x_{\textrm{in}},z_t)\gamma^*(x''+x_{\textrm{in}},z_t)  h_{\textrm{in}}(x',z_t,z_0,c_0) h^*_{\textrm{in}}(x',z_t,z_0,c_0) \Biggr \rangle \nonumber \\
 \propto \int dx' \int dx'' \, & h_{\textrm{out}}(x'-\Delta x,z_t,z_0,c_0)h^*_{\textrm{out}}(x''-\Delta x,z_t,z_0,c_0) \nonumber \\
 & \left \langle \gamma(x'+x_{\textrm{in}},z_t)\gamma^*(x''+x_{\textrm{in}},z_t) \right \rangle h_{\textrm{in}}(x',z_t,z_0,c_0) h^*_{\textrm{in}}(x',z_t,z_0,c_0). 
\label{incoherent}
\end{align}
Assuming a random speckle, 
\begin{equation}
\label{random}
\langle \gamma(x,z_t)  \gamma^*(x',z_t) \rangle =\langle |\gamma |^2\rangle \delta (x-x'),
\end{equation}
Eq.~\ref{incoherent} becomes:
\begin{align}
 RPSF^2_{\textrm{inc}}(\Delta x,c_0) \propto  & \int dx'   | h_{\textrm{out}}(x'-\Delta x,z_t,z_0,c_0)|^2 | h_{\textrm{in}}(x',z_t,z_0,c_0) |^2\\
& \propto |h_{\textrm{out}}|^2 \stackrel{\Delta x}{\circledast} |h_{\textrm{in}}|^2 (\Delta x,c_0). 
\label{incoherent2}
\end{align}
The incoherent RPSF therefore provides the auto-convolution of the input and output PSF intensities.

\section{Coherent RPSF}
\label{E}
A second option is to extract a coherent RPSF from the singular value decomposition (SVD) of $\mathbf{R}_{D}$. The result of the SVD can be understood if we assume, in first approximation, a point-like input PSF in Eq.~\ref{eqR0} [$h_{\textrm{in}}(x',z_t,z_0,c_0) \propto\delta(x')$]. Under this assumption, the $\mathbf{R}_D$-matrix coefficients can be expressed as follows:
\begin{equation}
\label{eqR}
R_D^{(1)}(\lbrace \Delta x,c_0 \rbrace,\lbrace x_\textrm{in},t \rbrace)= h_{\textrm{out}}(-\Delta x,c_0) \gamma(x_{\textrm{in}},z_t)
\end{equation}
where the superscript $(1)$ stands for the first-order approximation under which this expression has been derived. In this ideal case, Eq.~\ref{eqR} indicates that the $\mathbf{R}_D$-matrix is of rank 1. The corresponding eigenstate then directly provides the output PSF in the de-scanned basis, ${U}^{(1)}_1(\Delta x,c_0)=h_{\textrm{out}}(-\Delta x,c_0)$, and the phase conjugate of the medium reflectivity in the pixel basis, ${{V}^{(1)}_1(x_{\textrm{in}},t)}=\gamma^*(x_{\textrm{in}},\corr{z_t})$. 

These expressions result from a first-order approximation but can be improved using the relation that links the two singular vectors: $\lambda_1 \mathbf{U_1}=\mathbf{R}_D \times \mathbf{V_1}$. A second-order estimation of $\mathbf{U}_1$ can therefore be obtained by considering matrix product between the exact matrix $\mathbf{R}_D$ and the {first}-order estimation of $\mathbf{V}_1$, such that:
\begin{equation}
\label{eqU1}
{U_1}(\Delta x,c_0) \propto \sum_{\lbrace x_{\textrm{in}},t \rbrace } R_D(\lbrace \Delta x,c_0 \rbrace,\lbrace x_\textrm{in},t \rbrace) \gamma^{\corr{*}}(x_{\textrm{in}},z_t).
\end{equation}
Injecting Eq.~\ref{eqR0} into the last equation leads to 
\begin{equation}
\label{eqU1bis}
{U_1}(\Delta x,c_0) \propto \sum_{\lbrace x_{\textrm{in}},t \rbrace } \int dx' \, h_{\textrm{out}}(x'-\Delta x,z_t,z_0,c_0) h_{\textrm{in}}(x',z_t,z_0,c_0) \gamma(x'+x_{\textrm{in}},z_t)  \gamma^*(x_{\textrm{in}},z_t).
\end{equation}
If the number of {resolution cells} in the considered spatial window is sufficiently large, the sum over ${\lbrace x_{\textrm{in}},t \rbrace }$ in the last equation can be replaced by an ensemble average, such that
\begin{equation}
\label{eqU1tri}
{U_1}(\Delta x,c_0) \propto \int dx' \, h_{\textrm{out}}(x'-\Delta x,z_t,z_0,c_0) h_{\textrm{in}}(x',z_t,z_0,c_0) \langle \gamma(x'+x_{\textrm{in}},z_t)  \gamma^*(x_{\textrm{in}},z_t) \rangle ,
\end{equation}
where the symbol $\langle \cdots \rangle$ stands for the ensemble average. Assuming a random speckle ($\langle \gamma(x,z_t)  \gamma^*(x',z_t) \rangle =\langle |\gamma |^2\rangle \delta (x-x')$), Eq.~\ref{eqU1tri} finally yields to the following expression of $\mathbf{U}_1$:
\begin{equation}
\label{eqU1quad}
{U_1}(\Delta x,c_0) \propto  h_{\textrm{out}}(-\Delta x,z_t,z_0,c_0) h_{\textrm{in}}(0,z_t,z_0,c_0) . 
\end{equation}
The first singular vector therefore yields the amplitude distribution of the output focal spot, $h_{\textrm{out}}(-\Delta x,z_t,z_0,c_0)$, weighted by the confocal value of the input focal spot $h_{\textrm{in}}(0,z_t,z_0,c_0) $.

\section{Analytical expressions of the RPSFs for a Gaussian aperture function}
\label{F}

For analytical tractability, a Gaussian aperture function can be assumed in the expression of the transfer function $H(\kappa,z,\omega,c_0)$ (Eq.~\ref{approxH}) such that:
\begin{equation}
O_{\textrm{in/out}}(\kappa,z)=\exp \left [-\kappa^2/ (2A^2) \right ]
\end{equation}
For this Gaussian aperture, the resulting PSF $h$ (Eq.~\ref{impulse}) is a Gaussian beam:
\begin{equation}
\label{gaussian}
h_{\textrm{in/out}}(x,c_0)=\sqrt{\frac{w_0}{w(c_0)}}\exp \left (-\frac{x^2}{w^2(c_0)} -\corr{i}\frac{\omega_c}{c}z -\corr{i} \frac{x^2}{R^2(c_0)}+\corr{i}\eta(c_0)\right )
\end{equation}
with 
\begin{equation}
w(c_0)=w_0\sqrt{1+\frac{\left (c^2-c_0^2 \right )^2}{v^4}},
\end{equation}
the width of the Gaussian beam,
\begin{equation}
R(c_0)=w_0 \sqrt{\frac{v^2}{c^2-c_0^2}+ \frac{c^2-c_0^2}{v^2} },
\end{equation}
its radius of curvature, and
\begin{equation}
\eta(c_0)=\frac{1}{2}\arctan \left [\frac{c^2-c_0^2}{v^2}\right],
\end{equation}
the Gouy phase which implies a phase jump of $\pi/2$ in a 2D configuration. The beam waist, $w_0=\sqrt{2}/A$, and the characteristic velocity $v$ are related as follows: 
\begin{equation}
\label{v}
v^2=\frac{w_0^2 \omega_c}{t}.
\end{equation}

If we assume the input/output PSFs of Eq.~\ref{eqU1quad} as Gaussian beams, the incoherent RPSF (Eq.~\ref{incoherent2}) can be expressed as follows:
\begin{equation}
 RPSF_{\textrm{inc}}(\Delta x,c_0 ) \propto    \left ( \frac{w_0}{w(c_0)} \right)^{3/4} \exp \left (-\frac{\Delta x^2}{2w^2(c_0)} \right )  . 
\label{incoherent3}
\end{equation}
As to the coherent RPSF (Eq.~\ref{eqU1quad}), the Gaussian approximation leads to the following expression:
\begin{equation}
\label{eqU1cinq}
{RPSF}_{\textrm{coh}}(\Delta x,c_0) \propto \frac{w_0}{w(c_0)} \exp \left (-\frac{\Delta x^2}{w^2(c_0)} -\corr{i} \frac{\Delta x^2}{R^2(c_0)}+2\corr{i}\eta(c_0)\right )  . 
\end{equation}

\clearpage 

\section{Uncertainty}
\label{H}
To estimate the uncertainty of our speed-of-sound estimators, a Taylor development can be written around the RPSF maximum at $\Delta x=0$, such that
\begin{equation}
RPSF(c_0)=RPSF(c_0=c) + \frac{1}{2} (c-c_0)^2 \left (\frac{\partial^2 RPSF}{\partial c_0^2} \right )_{c_0=c}
\end{equation}
It leads to the following uncertainty relation:
\begin{equation}
\label{deltac}
\delta c  = \sqrt{\frac{ \delta RPSF}{ \left | \left ( \partial^2 RPSF/\partial c_0^2 \right )_{c_0=c} \right | }}
\end{equation}
with $\delta c$, the error of the estimator $\hat{c}$ and $\delta RPSF$, the standard deviation of the RPSF. The fluctuations of the RPSF can be expressed as follows
\begin{equation}
\frac{\delta RPSF_\textrm{inc}}{|RPSF_\textrm{inc}|} = \frac{1}{\beta\sqrt{N_{\mathcal{P}}}}
\end{equation}
with $\beta$, the signal-to-noise ratio and $N_{\mathcal{P}}$, the number of resolution cells contained in each spatial window $\mathcal{P}$ (Eq.~[4]). Injecting the last equation into Eq.~\ref{deltac} leads to the following expression for the error $\delta c_\textrm{inc}$ of a speed-of-sound estimator based on the incoherent RPSF: 
\begin{equation}
\label{deltac2}
\delta c  = \frac{1}{\beta^{1/2}{N_{\mathcal{P}}}^{1/4}} {\sqrt{ \frac{ |RPSF|}{ \left | \left ( \partial^2 RPSF/\partial c_0^2 \right )_{c_0=c} \right | }}}
\end{equation}

To go further, Gaussian PSFs are again assumed for sake of analytical tractability. If we first consider the incoherent RPSF (Eq.~\ref{incoherent3}), its second order derivative at $c_0=c$ is given by 
\begin{equation}
 \left ( \frac{\partial^2 RPSF_\textrm{inc}}{\partial c_0^2} \right )_{c_0=c}  = - 3\frac{c^2}{v^4} 
\end{equation}
Injecting this last result into Eq.~\ref{deltac2} and replacing $v^2$ by its expression (Eq.~\ref{v}) lead to the following uncertainty on $\hat{c}_\textrm{inc}$:
\begin{equation}
\label{deltacinc2}
\delta c_{\textrm{inc}}  = \frac{2}{\sqrt{3}}\frac{1}{\beta^{1/2}{N_{\mathcal{P}}}^{1/4} }\frac{z_R}{t}
\end{equation}
with $z_R=k w_0^2/2$, the Rayleigh range.

As to the modulus of the coherent RPSF,  
its second order derivative (Eq.~\ref{incoherent3}) at $c_0=c$ is given by 
\begin{equation}
 \left ( \frac{\partial^2 RPSF_\textrm{coh}}{\partial c_0^2} \right )_{c_0=c}  = - 4\frac{c^2}{v^4} RPSF_\textrm{coh}(c_0=c)
\end{equation}
Injecting this last result into Eq.~\ref{deltac2} and again replacing $v^2$ by its expression (Eq.~\ref{v}) lead to the following uncertainty on $\hat{c}_\textrm{coh}$:
\begin{equation}
\label{deltaccoh2}
\delta c_{\textrm{coh}}  =\frac{1}{\beta^{1/2}{N_{\mathcal{P}}}^{1/4} }\frac{z_R}{t}
\end{equation}
The wave velocity estimator based on the modulus of the coherent RPSF is therefore slightly better than the incoherent one since $\delta c_{\textrm{coh}}={\sqrt{3}}\delta c_{\textrm{inc}}/2 $.

Finally, the second order derivative of the real part of the coherent RPSF (Eq.~\ref{incoherent3}) at $c_0=c$ is given by 
\begin{equation}
 \left ( \frac{\partial^2 \mathcal{R} \lbrace RPSF_\textrm{coh} \rbrace }{\partial c_0^2} \right )_{c_0=c}  = - 20\frac{c^2}{v^4} RPSF_\textrm{coh}(c_0=c)
\end{equation}
Injecting this last result into Eq.~\ref{deltac2} leads to the following uncertainty on $\hat{c}_\textrm{gouy}$:
\begin{equation}
\label{deltacgouy2}
\delta c_{\textrm{gouy}}  = \frac{1}{\sqrt{5}}\frac{1}{\beta^{1/2}{N_{\mathcal{P}}}^{1/4} }\frac{z_R}{t}.
\end{equation}

\section{Effect of the numerical aperture}
\label{I}
\begin{figure}[htbp]
\includegraphics[width=\textwidth]{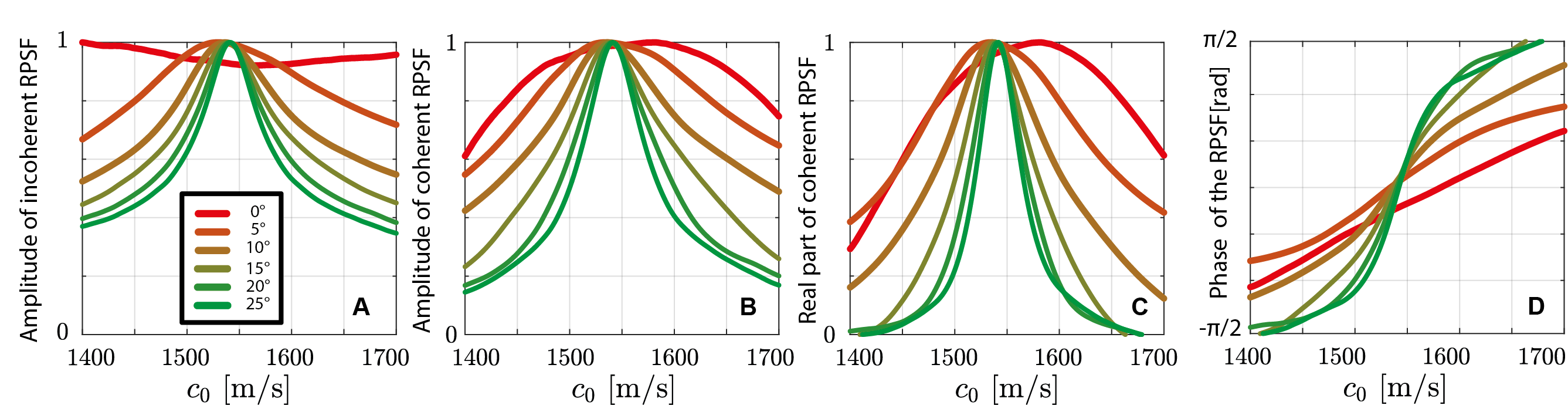}
    \caption{\textbf{Influence of the numerical aperture on the RPSFs.} ({A}) Amplitude of the confocal component of the incoherent RPSF versus $c_0$. ({B}) Amplitude of the confocal component of the coherent RPSF versus $c_0$. ({C}) Real part of the coherent RPSF versus $c_0$.  ({D}) Phase of the coherent RPSF versus $c_0$. Each observable is displayed for different numerical apertures from $\alpha=0^o$ (red) to {$\alpha=25^o$ (green)}. Results presented here correspond to the tissue mimicking phantom with parameters described in Table 1 of the accompanying paper. The selected point is located at $\mathbf{r}_{\textrm{p}}=(x_{\textrm{p}},t_{\textrm{p}})$ = ($0$ mm, $42.9$ $\mu$s) and the average window is $({p}_{x},{p}_{t}) = $ ($20$ mm, $2.6$ $\mu$s).}
    \label{figsupp_aperture}
\end{figure}

Equations~\ref{deltacinc2}, \ref{deltaccoh2} and \ref{deltacgouy2} show that the uncertainty $\delta c$ is directly proportional to the Rayleigh range $z_R\sim \lambda/NA^2$ and therefore decreases with the numerical aperture $NA=\sin \alpha$, with $\alpha $ the aperture angle. This effect is highlighted by Fig. \ref{figsupp_aperture} that shows the $c_0$-dependence of the incoherent RPSF (Fig. \ref{figsupp_aperture}A), the absolute value of the coherent RPSF  (Fig. \ref{figsupp_aperture}B), its real part (Fig. \ref{figsupp_aperture}C) and its phase (Fig. \ref{figsupp_aperture}D) along the focusing axis ($\Delta x=0$) and for different angular apertures. Not surprisingly, a sharper peak is observed for the three first curves around the optimized speed-of-sound value $c_p$ when the numerical aperture increases (Fig. \ref{figsupp_aperture}C). This effect is even more drastic on the real part of the RPSF since the Gouy phase jump also becomes steeper at large numerical apertures (Fig. \ref{figsupp_aperture}D). A sharper RPSF peak implies a larger second order derivative of the RPSF at its maximmal value and thus a lower uncertinty on our estimation of the speed-of-sound $c$ (Eq.~\ref{deltac2}). Figure \ref{figsupp_aperture} is therefore a striking illustration of the uncertainty reduction scaling as $NA^{-2}$.

\section{Numerical simulations}
\label{J}

\begin{table*}[!ht]
    \center
\begin{tabular}{ll|ll}
    &  \multicolumn{2}{l}{\textbf{Parameters}} & \multicolumn{1}{l}{\textbf{Value}}\\
      \hline
       \multirow{4}{*}{\textbf{Sampling}}
    & \multirow{2}{*}{{Spatial grid}}
        & Number of points $N_x \times N_z$  & $2000\times 2000$ \\ 
        & & Spatial sampling & $\lambda/10 \sim 50$ $\mu$m\\ 
        & \multicolumn{2}{l}{{Sampling frequency $f_s$}}  & $102$ MHz \\ 
    & \multicolumn{2}{l}{{Recording time}}  & $166$ $\mu$s \\ 
    \hline
    \multirow{3}{*}{\textbf{Medium}}
    & \multicolumn{2}{l}{{Speed-of-sound}} & Figs.~\ref{figsupp_simu}A,B \\
     & \multirow{2}{*}{{Density}}
        & mean $\langle\rho\rangle$ & {1000} kg.m$^{-3}$\\ 
        & & standard deviation  & {3} kg.m$^{-3}$\\ 
        \hline       
    \multirow{5}{*}{\textbf{Probe}}
    & \multicolumn{2}{l}{{Type}}  & Linear\\
    & \multicolumn{2}{l}{{Number of transducers}}  & $376$ \\ 
    & \multicolumn{2}{l}{{Transducer pitch $\delta u$}}  & $\lambda/2 \sim 0.26$ mm \\ 
    & \multicolumn{2}{l}{{Central frequency $f_c$}} & $3$ MHz\\
    & \multicolumn{2}{l}{{Bandwidth $\Delta f$}} & $[2-4]$ MHz\\
    \hline
    \multirow{4}{*}{\textbf{Acquisition}}
    & \multicolumn{2}{l}{{Speed-of-sound hypothesis $c_0$}} & $1540$ m/s \\
    & \multirow{3}{*}{{Plane wave angles}}
        &  maximum $\theta_\inm$ & $30^\circ$\\ 
       & & pitch $\delta \theta_\inm$ & $1^\circ$ \\ 
        & & number $N_{\theta_\inm}$ & $61$ \\
        \hline
\end{tabular}
\caption{\textbf{Parameters of the k-wave numerical simulations.} }
\label{AcquisitionParamSimu}
\end{table*}

Each k-Wave simulation is performed over a two-dimensional grid whose sampling is reported in Tab.~\ref{AcquisitionParamSimu}. The simulated speed-of-sound distributions are displayed in Figs.~\ref{figsupp_simu}A and B. The background density is constant ($\rho=1000$ kg.m$^{-3}$) but short-scale fluctuations (3 kg.m$^{-3}$ std) have been superimposed to generate a random wave-field characteristic of ultrasound imaging in soft tissues. {Bright targets are point-like threads modeled by a higher density $\rho_t=1100$  kg.m$^{-3}$ compared to \corr{background}.}

All simulation parameters such as the probe configuration and acquisition sequence are also described in Tab.~\ref{AcquisitionParamSimu}. As in the phantom experiment, the recording of the reflection matrix is performed using a set of incident plane waves. The emitted signal is a {3 MHz} two-cycle sinusoidal burst. For each excitation, the back-scattered signal is recorded by the probe and stored in the reflection matrix $\mathbf{R}_{\mathbf{u}{\theta}}(t)$. 

\clearpage

\section{Delay-and-sum algorithm - curved array}
\label{L}
For a curved array, the ultrasound image is expressed in polar coordinates $(\alpha,\rho)$. A set of diverging waves is generated by applying the same linear time delay that we would apply to generate a plane wave of incidence angle $\theta_{\textrm{in}}$ from a linear array. 
The beamforming process used to build the confocal image displayed in Fig.~\ref{fig_invivo_OptimisationImage}A can therefore be expressed as follows:
\begin{align}
\mathcal{I}(\alpha,r_0=c_0t/2)=\sum_{\theta_\inm}\sum_{u_\outm}&R(u_\outm,\theta_\textrm{in},\tau_{\textrm{out}}(\mathbf{u}_\outm,\alpha,t,c_0) \nonumber\\
&+\tau_{\textrm{in}}(\theta_\textrm{in},\alpha,t,c_0)).
\label{confocalcurved}
\end{align}
with $\mathbf{u}_\outm=(v_\outm,w_\outm)$, the position vector of each transducer. The input travel times are given by :
\begin{equation}
{\tau_{\textrm{in}}(\theta_\textrm{in},\alpha,t,c_0)=\frac{D+\mathcal{R}_u \corr{\theta_p}\sin(\corr{\theta_{\textrm{in}}^0})}{c_0}}
\end{equation}
with $\mathcal{R}_u$, the curvature radius of the probe and  
\begin{align}
&{D=\sqrt{\rho^{\prime2}-[\mathcal{R}_u\sin(\corr{\theta_{\textrm{in}}^0})]^2}}-\corr{\mathcal{R}_u \cos \theta_{\textrm{in}}^0} \\
&{\theta_p=\arcsin \lbrace [\sin(\alpha)(\mathcal{R}_u+D\cos(\corr{\theta_{\textrm{in}}^0}))-D\cos(\alpha)\sin(\corr{\theta_{\textrm{in}}^0})]/\rho^\prime\rbrace }\\
&{\rho^\prime=\mathcal{R}_u+(c_\textrm{acq}t/2-\mathcal{R}_u)\frac{c_0}{c_\textrm{acq}}}
\end{align}
The output travel times are given by :
\begin{equation}
{\tau_{\textrm{out}}(\mathbf{u}_\outm,\alpha,t,c_0)=\frac{\sqrt{\left[\rho^\prime\sin(\alpha)-v_\outm\right]^2+\left[\rho^\prime\cos(\alpha)-w_\outm\right]^2}}{c_0}}
\end{equation}
These travel times are also used to build the focused reflection matrix, $\mathbf{R}_{\alpha \alpha}(t,c_0)$ whose coefficients are given by:
\begin{align}
R(\alpha_\textrm{out},\alpha_\textrm{in},t,c_0)=\sum_{\theta_\inm}\sum_{u_\outm}&R(\mathbf{u}_\outm,\theta_\textrm{in}, \tau_\textrm{out}(\mathbf{u}_\outm,\alpha_\textrm{out},t,c_0)\nonumber\\
&+\tau_\textrm{in}(\theta_\textrm{in},\alpha_\textrm{in},t,c_0)).
\label{focusedR2}
\end{align}
One example of the matrix $\mathbf{R}_{\alpha \alpha}(t,c_0)$ is given in Fig.~\ref{fig_invivo_OptimisationImage}E.

\section{Optimal and local speed-of-sound}
\label{N}

The optimal speed-of-sound $\hat{c}(\mathbf{r}_{\textrm{p}})$ is estimated locally by considering the de-scanned matrix over a reduced spatial window $\mathcal{P}(\mathbf{r}-\mathbf{r}_{\textrm{p}})$ (Eq.~\ref{P}). The choice of this window is dictated by the following compromise: (\textit{i}) encompass a sufficient number of independent speckle points to smooth out speckle fluctuations in the RPSF by spatial averaging; (\textit{ii}) consider a spatial window as small as possible in order to optimize the resolution of the $\hat{c}$-map. The size of $\mathcal{P}$ resulting from this compromise is reported in {Tab.~\ref{LiverAverageTablesmooth}} for the experiments and numerical simulations shown in the paper. 

The optimal speed-of-sound map, $\hat{c} (x,t)$, can finally be used to estimate a local speed-of-sound map, ${c}(x,z)$. To this end, we take up the method developed by Jakovljevic \textit{et al.}~\cite{jakovljevic_local_2018} which basically consists in differentiating the first order eikonal equation. 

Under a paraxial approximation and assuming that the speed of sound, or equivalently the slowness $s=1/c$, is a piecewise constant function between discretized depths, such as $z_n=n\delta z$ with $n\in\{0,1,...,N\}$, we can write:
\begin{align}
    \underbrace{\int_{z_{n-1}}^{z_n}s(x,z)\textrm{d}z}_{\approx {s(x,z_n)\delta z}}=&\int_{0}^{z_n}s(x,z)\textrm{d}z-\int_{0}^{z_{n-1}}s(x,z)\textrm{d}z;
\end{align}
which yields
\begin{equation}
{s(x,z_n)\delta z}= n\bar{s}(x,z_n)\delta z-(n-1)\bar{s}(x,z_{n-1})\delta z
\end{equation}
with $\bar{s}(x,z)$, the depth-averaged slowness from the probe to the depth $z$. The last equation leads to a discretized expression of the local slowness $s$ as a numerical differentiation of the averaged slowness $\bar{s}$:
\begin{equation}
  s(x,z_n)= -(n-1)\bar{s}(x,z_{n-1})+n\bar{s}(x,z_n). \label{differentiationsos}
\end{equation}
The last equation leads to the following system of equations:
\begin{align}
\left\{
\begin{array}{ll}
        s(x,z_1)=& \bar{s}(x,z_1); \\
        s(x,z_2)=& -\bar{s}(x,z_1)+2\bar{s}(x,z_2); \\
        s(x,z_3)=& -2\bar{s}(x,z_2)+3\bar{s}(x,z_3);\\
        ...\\
        s(x,z_N)=& -(N-1)\bar{s}(x,z_{N-1})+N\bar{s}(x,z_N).
    \end{array}
    \right.
    \label{systemspeedofsoundFlav}
\end{align}
Under a matrix formalism, it writes: 
\begin{equation} 
\mathbf{S}(x)= \mathbf{A} \times \bar{\mathbf{S}}(x).
\label{sosInversionAppendix}
\end{equation}
where $\mathbf{S}(x)=[s(x,z_i)]^\top_i$ and $\bar{\mathbf{S}}(x)=[\bar{s}(x,z_i)]^\top_i$ are column vectors containing the discretized values of the local slowness $s(x,z)=1/c(x,z)$ and of the depth-averaged slowness $\bar{s}(x,z_t)=1/\hat{c}(x,t)$ interpolated at each depth $z_i$. $\mathbf{A}$ is the following matrix : 
 \begin{equation}
\mathbf{A}=\begin{pmatrix}
1 & 0 & 0  &...& 0 & 0\\
-1 & 2 & 0  & ...& 0 & 0\\
0 & -2 & 3  & ...& 0 & 0\\
... & ... & ... &...& ...  & ...\\
0 & 0 & 0 & ... & -(N-1) & N\\
\end{pmatrix}.
\label{EquGradient1}
\end{equation}
Equation~\ref{sosInversionAppendix} shows how the local velocity $c(x,z)=1/s(x,z)$ can be obtained from a numerical differentiation of the depth-averaged slowness $\bar{s}(x,z_t)=1/\hat{c}(x,t)$. 

\corr{This numerical differentiation is, however, unstable since our measurement $\hat{c}(x,t)$ is invariably corrupted by noise.} A regularization method is therefore needed and consists in a prior smoothing of the averaged speed-of-sound in order to avoid un-physical short-scale fluctuations of the local speed-of-sound. In practice, the smoothing operation is performed in two steps.

First, the RPSF is smoothed by means of a spatio-temporal Gaussian kernel such that:
\begin{equation}
   RPSF_{\textrm{inc}}(\boldsymbol{\Delta},\mathbf{r}_{\textrm{p}})={\sqrt{\left\lvert R(\boldsymbol{\Delta},\mathbf{r}_{\textrm{p}})\right\rvert^2 \overset{\mathbf{r}_{\textrm{p}}}{\circledast} \mathcal{K}(\mathbf{r}_{\textrm{p}})}},
\end{equation}
where $\mathcal{K}$ is a Gaussian kernel  such that $\mathcal{K}(\mathbf{r})=\mathcal{K}(x,t)=\exp[-x^2/(2l_x^2)]\exp[-t^2/(2l_t^2)]$ for the numerical simulation (Fig.~\ref{figsupp_simu}, linear array) and $\mathcal{K}(\mathbf{r})=\mathcal{K}(\alpha,t)=\exp[-\alpha^2/(2l_{\alpha}^2)]\exp[-t^2/(2l_t^2)]$ for the liver experiment (Fig.~\ref{fig_invivo_SOS}, curved array). The values of $l_x$, $l_{\alpha}$ and $l_t$ are provided in Tab.~\ref{LiverAverageTablesmooth}.

Second, the $\hat{c}$-map, extracted from the maximization of $RPSF_{\textrm{inc}}$, is smoothed using an equivalent Gaussian kernel $\mathcal{K}$, whose dimension $(l_{x/\alpha},l_t)$ is also provided in Tab.~\ref{LiverAverageTablesmooth}.  

\begin{table}[!ht]
    \center
\begin{tabular}{cccc}
   \multicolumn{1}{c}{\textbf{Quantity}}  & \multicolumn{1}{c}{\textbf{Direction}} & \multicolumn{1}{c}{\textbf{Simulations}} &  \multicolumn{1}{c}{\textbf{Liver in-vivo}} \\
      \midrule
    \multirow{2}{*}{RPSF}  &  {Lateral}  &  $l_x=2.5$ mm ($\sim 5 \lambda$) &  $l_\alpha =4.5$ $^\circ$ \\ 
    &    {Axial} & $l_t=1.6$ $\mu$s & $l_t=3.2$ $\mu$s   \\ 
    \midrule
     \multirow{2}{*}{$\hat{c}$-map }  &  {Lateral}  & $l_x=2.5$ mm ($\sim 5 \lambda$) & $l_\alpha =2.5$ $^\circ$ \\ 
    &    {Axial} & $l_t=1.6$ $\mu$s & $l_t=1.0$ $\mu$s  \\ 
    \bottomrule
\end{tabular}
\caption{Size of the Gaussian kernels used for smoothing the RPSF and $\hat{c}$-map prior to the numerical differentiation of Eq.~\ref{sosInversionAppendix}.}
\label{LiverAverageTablesmooth}
\end{table}

\clearpage

\bibliography{references2}

@Article{bureau_three-dimensional_2023,
  author  = {Bureau, Flavien and Robin, Justine and Le Ber, Arthur and Lambert, William and Fink, Mathias and Aubry, Alexandre},
  title   = {Three-dimensional ultrasound matrix imaging},
  doi     = {https://doi.org/10.1038/s41467-023-42338-8},
  pages   = {6793},
  volume  = {14},
  journal = {Nat. Commun.},
  year    = {2023},
}

@Article{chen_multi-layer_2020,
  author   = {Chen, M. and Ren, D. and Liu, H.-Y. and Chowdhury, S. and Waller, L.},
  title    = {Multi-layer {Born} multiple-scattering model for {3D} phase microscopy},
  doi      = {10.1364/OPTICA.383030},
  pages    = {394},
  urldate  = {2023-10-12},
  volume   = {7},
  journal  = {Optica},
  month    = may,
  year     = {2020},
}

@Article{babcock_possibility_1953,
  author  = {Babcock, H. W.},
  title   = {The possibility of compensating astronomical seeing},
  doi     = {10.1086/126606},
  issn    = {0004-6280},
  number  = {386},
  pages   = {229--236},
  url     = {https://www.jstor.org/stable/40672682},
  urldate = {2022-04-04},
  volume  = {65},
  journal = {Publ. Astron. Soc. Pac.},
  year    = {1953},
}

@Article{mehta_non-invasive_2008,
  author  = {Mehta, Sanjeev R and Thomas, E Louise and Bell, Jimmy D and Johnston, Desmond G and Taylor-Robinson, Simon D},
  title   = {Non-invasive means of measuring hepatic fat content},
  doi     = {10.3748/wjg.14.3476},
  issn    = {1007-9327},
  number  = {22},
  pages   = {3476},
  url     = {http://www.wjgnet.com/1007-9327/full/v14/i22/3476.htm},
  urldate = {2023-05-15},
  volume  = {14},
  journal = {World J. Gastroenterol.},
  year    = {2008},
}

@Article{dasarathy_validity_2009,
  author     = {Dasarathy, Srinivasan and Dasarathy, Jaividhya and Khiyami, Amer and Joseph, Rajesh and Lopez, Rocio and McCullough, Arthur J.},
  title      = {Validity of real time ultrasound in the diagnosis of hepatic steatosis: {A} prospective study},
  doi        = {10.1016/j.jhep.2009.09.001},
  issn       = {01688278},
  number     = {6},
  pages      = {1061--1067},
  url        = {https://linkinghub.elsevier.com/retrieve/pii/S0168827809005856},
  urldate    = {2023-05-15},
  volume     = {51},
  journal    = {J. Hepatol.},
  month      = dec,
  shorttitle = {Validity of real time ultrasound in the diagnosis of hepatic steatosis},
  year       = {2009},
}

@Article{Giraudat2025b,
  author  = {Giraudat, E. and Bureau, F. and Lambert, W. and Fink, M. and Aubry, A.},
  title   = {Self-Portrait of the Focusing Process in Speckle: {III. T}ailoring Complex Spatio-Temporal Focusing Laws To Overcome Reverberations in Reflection Imaging},
  journal = {arXiv},
  year    = {2026},
}

@Article{heriard-dubreuil_refraction-based_2023,
  author     = {Hériard-Dubreuil, Baptiste and Besson, Adrien and Wintzenrieth, Frédéric and Cohen-Bacrie, Claude and Thiran, Jean-Philippe},
  title      = {Refraction-{Based} {Speed} of {Sound} {Estimation} in {Layered} {Media}: an {Angular} {Approach}},
  doi        = {10.1109/TUFFC.2023.3261541},
  issn       = {0885-3010, 1525-8955},
  pages      = {486--497},
  url        = {https://ieeexplore.ieee.org/document/10081032/},
  urldate    = {2023-04-14},
  volume     = {70},
  journal    = {IEEE Trans. Ultrason. Ferroel. Freq. Cont.},
  shorttitle = {Refraction-{Based} {Speed} of {Sound} {Estimation} in {Layered} {Media}},
  year       = {2023},
}

@Article{hirama_adaptive_1982,
  author  = {Hirama, Makoto and Ikeda, Osamu and Sato, Takuso},
  title   = {Adaptive ultrasonic array imaging system through an inhomogeneous layer},
  doi     = {10.1121/1.387336},
  issn    = {0001-4966},
  number  = {1},
  pages   = {100--109},
  url     = {https://pubs.aip.org/asa/jasa/article/71/1/100-109/763204},
  urldate = {2023-05-09},
  volume  = {71},
  journal = {J. Acoust. Soc. Am.},
  month   = jan,
  year    = {1982},
}

@Article{ozmen_comparing_2015,
  author   = {Ozmen, Neslihan and Dapp, Robin and Zapf, Michael and Gemmeke, Hartmut and Ruiter, Nicole V. and Van Dongen, Koen W. A.},
  title    = {Comparing different ultrasound imaging methods for breast cancer detection},
  doi      = {10.1109/TUFFC.2014.006707},
  issn     = {0885-3010},
  number   = {4},
  pages    = {637--646},
  url      = {http://ieeexplore.ieee.org/document/7081460/},
  urldate  = {2023-05-05},
  volume   = {62},
  journal  = {IEEE Trans. Ultrason. Ferroel. Freq. Cont.},
  month    = apr,
  year     = {2015},
}

@Article{ruby_breast_2019,
  author     = {Ruby, Lisa and Sanabria, Sergio J. and Martini, Katharina and Dedes, Konstantin J. and Vorburger, Denise and Oezkan, Ece and Frauenfelder, Thomas and Goksel, Orcun and Rominger, Marga B.},
  title      = {Breast {Cancer} {Assessment} {With} {Pulse}-{Echo} {Speed} of {Sound} {Ultrasound} {From} {Intrinsic} {Tissue} {Reflections}: {Proof}-of-{Concept}},
  doi        = {10.1097/RLI.0000000000000553},
  issn       = {0020-9996},
  number     = {7},
  pages      = {419--427},
  url        = {http://journals.lww.com/00004424-201907000-00005},
  urldate    = {2023-05-05},
  volume     = {54},
  journal    = {Invest. Radiol.},
  month      = jul,
  shorttitle = {Breast {Cancer} {Assessment} {With} {Pulse}-{Echo} {Speed} of {Sound} {Ultrasound} {From} {Intrinsic} {Tissue} {Reflections}},
  year       = {2019},
}

@InCollection{duck_acoustic_1990,
  author    = {Duck, Francis A.},
  booktitle = {Physical {Properties} of {Tissues}},
  title     = {Acoustic {Properties} of {Tissue} at {Ultrasonic} {Frequencies}},
  doi       = {10.1016/B978-0-12-222800-1.50008-5},
  isbn      = {978-0-12-222800-1},
  pages     = {73--135},
  publisher = {Elsevier},
  url       = {https://linkinghub.elsevier.com/retrieve/pii/B9780122228001500085},
  urldate   = {2023-05-05},
  year      = {1990},
}

@Article{treeby_k-wave_2010,
  author     = {Treeby, Bradley E. and Cox, B. T.},
  title      = {k-{Wave}: {MATLAB} toolbox for the simulation and reconstruction of photoacoustic wave fields},
  doi        = {10.1117/1.3360308},
  issn       = {10833668},
  number     = {2},
  pages      = {021314},
  url        = {http://biomedicaloptics.spiedigitallibrary.org/article.aspx?doi=10.1117/1.3360308},
  urldate    = {2023-05-05},
  volume     = {15},
  journal    = {J. Biomed. Opt.},
  shorttitle = {k-{Wave}},
  year       = {2010},
}

@Article{montaldo_time_2011,
  author   = {Montaldo, Gabriel and Tanter, Mickael and Fink, Mathias},
  title    = {Time {Reversal} of {Speckle} {Noise}},
  doi      = {10.1103/PhysRevLett.106.054301},
  number   = {5},
  pages    = {054301},
  url      = {https://link.aps.org/doi/10.1103/PhysRevLett.106.054301},
  urldate  = {2022-04-04},
  volume   = {106},
  journal  = {Phys. Rev. Lett.},
  month    = feb,
  year     = {2011},
}

@Article{masoy_iteration_2005,
  author  = {Måsøy, Svein-Erik and Varslot, Trond and Angelsen, Bjørn},
  title   = {Iteration of transmit-beam aberration correction in medical ultrasound imaging},
  doi     = {10.1121/1.1823213},
  issn    = {0001-4966},
  number  = {1},
  pages   = {450--461},
  url     = {http://asa.scitation.org/doi/10.1121/1.1823213},
  urldate = {2022-11-09},
  volume  = {117},
  journal = {J. Acoust. Soc. Am.},
  month   = jan,
  year    = {2005},
}

@InProceedings{beuret_refraction-aware_2020,
  author    = {Beuret, Samuel and Perdios, Dimitris and Thiran, Jean-Philippe},
  booktitle = {2020 {IEEE} {International} {Ultrasonics} {Symposium} ({IUS})},
  title     = {Refraction-{Aware} {Integral} {Operator} for {Speed}-of-{Sound} {Pulse}-{Echo} {Imaging}},
  doi       = {10.1109/IUS46767.2020.9251601},
  pages     = {1--4},
  month     = sep,
  year      = {2020},
}

@InProceedings{dahl_phase_2005,
  author    = {Dahl, J.J. and Ivancevich, N.M. and Keen, C.G. and Trahey, G.E. and Smith, S.W.},
  booktitle = {{IEEE} {Ultrasonics} {Symposium}, 2005.},
  title     = {Phase correction of skull aberration with 1.75-{D} and 2-{D} arrays using speckle targets},
  doi       = {10.1109/ULTSYM.2005.1603097},
  isbn      = {978-0-7803-9382-0},
  pages     = {1323--1326},
  publisher = {IEEE},
  url       = {http://ieeexplore.ieee.org/document/1603097/},
  urldate   = {2023-03-17},
  volume    = {2},
  address   = {Rotterdam, The Netherlands},
  year      = {2005},
}

@Article{ali_distributed_2022,
  author   = {Ali, Rehman and Brevett, Thurston and Hyun, Dongwoon and Brickson, Leandra L. and Dahl, Jeremy J.},
  title    = {Distributed {Aberration} {Correction} {Techniques} {Based} on {Tomographic} {Sound} {Speed} {Estimates}},
  doi      = {10.1109/TUFFC.2022.3162836},
  issn     = {0885-3010, 1525-8955},
  number   = {5},
  pages    = {1714--1726},
  url      = {https://ieeexplore.ieee.org/document/9745035/},
  urldate  = {2022-11-09},
  volume   = {69},
  journal  = {IEEE Transactions on Ultrasonics, Ferroelectrics, and Frequency Control},
  month    = may,
  year     = {2022},
}

@Article{yoon_optimal_2012,
  author   = {Yoon, Changhan and Seo, Haijin and Lee, Yuhwa and Yoo, Yangmo and Song, Tai-Kyong and Chang, Jin Ho},
  title    = {Optimal sound speed estimation using modified nonlinear anisotropic diffusion to improve spatial resolution in ultrasound imaging},
  doi      = {10.1109/TUFFC.2012.2275},
  issn     = {1525-8955},
  number   = {5},
  pages    = {905--914},
  volume   = {59},
  journal  = {IEEE Trans. Ultrason. Ferroel. Freq. Cont.},
  month    = may,
  pmid     = {22622975},
  year     = {2012},
}

@Article{perrot_so_2021,
  author     = {Perrot, Vincent and Polichetti, Maxime and Varray, François and Garcia, Damien},
  title      = {So you think you can {DAS}? {A} viewpoint on delay-and-sum beamforming},
  doi        = {10.1016/j.ultras.2020.106309},
  issn       = {0041-624X},
  pages      = {106309},
  url        = {https://www.sciencedirect.com/science/article/pii/S0041624X20302444},
  urldate    = {2022-10-14},
  volume     = {111},
  journal    = {Ultrasonics},
  month      = mar,
  shorttitle = {So you think you can {DAS}?},
  year       = {2021},
}

@Article{benjamin_novel_2018,
  author   = {Benjamin, Alex and Zubajlo, Rebecca E. and Dhyani, Manish and Samir, Anthony E. and Thomenius, Kai E. and Grajo, Joseph R. and Anthony, Brian W.},
  title    = {A {Novel} {Approach} to the {Quantification} of the {Longitudinal} {Speed} of {Sound} and its {Potential} for {Tissue} {Characterization} ({Part} - {I})},
  doi      = {10.1016/j.ultrasmedbio.2018.07.021},
  issn     = {0301-5629},
  number   = {12},
  pages    = {2739--2748},
  url      = {https://www.ncbi.nlm.nih.gov/pmc/articles/PMC6662181/},
  urldate  = {2022-04-08},
  volume   = {44},
  journal  = {Ultrasound Med. Biol.},
  month    = dec,
  pmcid    = {PMC6662181},
  pmid     = {30228044},
  year     = {2018},
}

@Article{yoon_vitro_2011,
  author   = {Yoon, Changhan and Lee, Yuhwa and Chang, Jin Ho and Song, Tai-Kyong and Yoo, Yangmo},
  title    = {In vitro estimation of mean sound speed based on minimum average phase variance in medical ultrasound imaging},
  doi      = {10.1016/j.ultras.2011.03.007},
  issn     = {1874-9968},
  number   = {7},
  pages    = {795--802},
  volume   = {51},
  journal  = {Ultrasonics},
  month    = oct,
  pmid     = {21459400},
  year     = {2011},
}

@Article{zubajlo_experimental_2018,
  author   = {Zubajlo, Rebecca E. and Benjamin, Alex and Grajo, Joseph R. and Kaliannan, Kanakaraju and Kang, Jing X. and Bhan, Atul K. and Thomenius, Kai E. and Anthony, Brian W. and Dhyani, Manish and Samir, Anthony E.},
  title    = {Experimental {Validation} of {Longitudinal} {Speed} of {Sound} {Estimates} in the {Diagnosis} of {Hepatic} {Steatosis} ({Part} {II})},
  doi      = {10.1016/j.ultrasmedbio.2018.07.020},
  issn     = {1879-291X},
  number   = {12},
  pages    = {2749--2758},
  volume   = {44},
  journal  = {Ultrasound Med. Biol.},
  month    = dec,
  pmcid    = {PMC6661157},
  pmid     = {30266215},
  year     = {2018},
}

@Article{feng_physical_2001,
  author  = {Feng, Simin and Winful, Herbert G.},
  title   = {Physical origin of the {Gouy} phase shift},
  doi     = {10.1364/OL.26.000485},
  issn    = {0146-9592, 1539-4794},
  number  = {8},
  pages   = {485},
  url     = {https://opg.optica.org/abstract.cfm?URI=ol-26-8-485},
  urldate = {2022-06-03},
  volume  = {26},
  journal = {Opt. Lett.},
  month   = apr,
  year    = {2001},
}

@Article{stahli_improved_2020,
  author   = {Stähli, Patrick and Kuriakose, Maju and Frenz, Martin and Jaeger, Michael},
  title    = {Improved forward model for quantitative pulse-echo speed-of-sound imaging},
  doi      = {10.1016/j.ultras.2020.106168},
  issn     = {0041624X},
  pages    = {106168},
  url      = {https://linkinghub.elsevier.com/retrieve/pii/S0041624X20301074},
  urldate  = {2021-11-03},
  volume   = {108},
  journal  = {Ultrasonics},
  month    = dec,
  year     = {2020},
}

@Article{anderson_direct_1998,
  author  = {Anderson, Martin E. and Trahey, Gregg E.},
  title   = {The direct estimation of sound speed using pulse–echo ultrasound},
  doi     = {10.1121/1.423889},
  issn    = {0001-4966},
  number  = {5},
  pages   = {3099--3106},
  url     = {https://asa.scitation.org/doi/10.1121/1.423889},
  urldate = {2022-04-08},
  volume  = {104},
  journal = {J. Acoust. Soc. Am.},
  month   = nov,
  year    = {1998},
}

@Article{jakovljevic_local_2018,
  author     = {Jakovljevic, Marko and Hsieh, Scott and Ali, Rehman and Chau Loo Kung, Gustavo and Hyun, Dongwoon and Dahl, Jeremy J.},
  title      = {Local speed of sound estimation in tissue using pulse-echo ultrasound: {Model}-based approach},
  doi        = {10.1121/1.5043402},
  issn       = {0001-4966},
  number     = {1},
  pages      = {254--266},
  url        = {https://www.ncbi.nlm.nih.gov/pmc/articles/PMC6045494/},
  urldate    = {2022-04-08},
  volume     = {144},
  journal    = {J. Acoust. Soc. Am.},
  month      = jul,
  pmcid      = {PMC6045494},
  pmid       = {30075660},
  shorttitle = {Local speed of sound estimation in tissue using pulse-echo ultrasound},
  year       = {2018},
}

@Article{imbault_robust_2017,
  author  = {Imbault, Marion and Faccinetto, Alex and Osmanski, Bruno-Félix and Tissier, Antoine and Deffieux, Thomas and Gennisson, Jean-Luc and Vilgrain, Valérie and Tanter, Mickaël},
  title   = {Robust sound speed estimation for ultrasound-based hepatic steatosis assessment},
  doi     = {10.1088/1361-6560/aa6226},
  issn    = {0031-9155, 1361-6560},
  number  = {9},
  pages   = {3582--3598},
  url     = {https://iopscience.iop.org/article/10.1088/1361-6560/aa6226},
  urldate = {2021-11-02},
  volume  = {62},
  journal = {Phys. Med. Biol.},
  month   = may,
  year    = {2017},
}

@Article{odonnell_phase-aberration_1988,
  author     = {O'Donnell, M. and Flax, S.W.},
  title      = {Phase-aberration correction using signals from point reflectors and diffuse scatterers: measurements},
  doi        = {10.1109/58.9334},
  issn       = {0885-3010},
  number     = {6},
  pages      = {768--774},
  url        = {http://ieeexplore.ieee.org/document/9334/},
  urldate    = {2022-04-12},
  volume     = {35},
  journal    = {IEEE Trans. Ultras. Ferroel. Freq. Cont.},
  month      = nov,
  shorttitle = {Phase-aberration correction using signals from point reflectors and diffuse scatterers},
  year       = {1988},
}

@Article{napolitano_sound_2006,
  author   = {Napolitano, David and Chou, Ching-Hua and McLaughlin, Glen and Ji, Ting-Lan and Mo, Larry and DeBusschere, Derek and Steins, Robert},
  title    = {Sound speed correction in ultrasound imaging},
  doi      = {10.1016/j.ultras.2006.06.061},
  issn     = {0041-624X},
  pages    = {e43--e46},
  series   = {Proceedings of {Ultrasonics} {International} ({UI}’05) and {World} {Congress} on {Ultrasonics} ({WCU})},
  url      = {https://www.sciencedirect.com/science/article/pii/S0041624X06002976},
  urldate  = {2022-04-12},
  volume   = {44},
  journal  = {Ultrasonics},
  month    = dec,
  year     = {2006},
}

@Article{hinkelman_measurements_1997,
  author  = {Hinkelman, Laura M. and Szabo, Thomas L. and Waag, Robert C.},
  title   = {Measurements of ultrasonic pulse distortion produced by human chest wall},
  doi     = {10.1121/1.418248},
  issn    = {0001-4966},
  number  = {4},
  pages   = {2365--2373},
  url     = {http://asa.scitation.org/doi/10.1121/1.418248},
  urldate = {2022-04-12},
  volume  = {101},
  journal = {J. Acoust. Soc. Am.},
  month   = apr,
  year    = {1997},
}

@Article{shin_estimation_2010,
  author   = {Shin, Ho-Chul and Prager, Richard and Gomersall, Henry and Kingsbury, Nick and Treece, Graham and Gee, Andrew},
  title    = {Estimation of speed of sound in dual-layered media using medical ultrasound image deconvolution},
  doi      = {10.1016/j.ultras.2010.02.008},
  issn     = {1874-9968},
  number   = {7},
  pages    = {716--725},
  volume   = {50},
  journal  = {Ultrasonics},
  month    = jun,
  pmid     = {20231026},
  year     = {2010},
}

@Article{jaeger_computed_2015,
  author     = {Jaeger, Michael and Held, Gerrit and Peeters, Sara and Preisser, Stefan and Grünig, Michael and Frenz, Martin},
  title      = {Computed ultrasound tomography in echo mode for imaging speed of sound using pulse-echo sonography: proof of principle},
  doi        = {10.1016/j.ultrasmedbio.2014.05.019},
  issn       = {1879-291X},
  number     = {1},
  pages      = {235--250},
  volume     = {41},
  journal    = {Ultrasound Med. Biol.},
  month      = jan,
  pmid       = {25220274},
  shorttitle = {Computed ultrasound tomography in echo mode for imaging speed of sound using pulse-echo sonography},
  year       = {2015},
}

@Article{montaldo_coherent_2009,
  author   = {Montaldo, Gabriel and Tanter, Mickaël and Bercoff, Jérémy and Benech, Nicolas and Fink, Mathias},
  title    = {Coherent plane-wave compounding for very high frame rate ultrasonography and transient elastography},
  doi      = {10.1109/TUFFC.2009.1067},
  issn     = {1525-8955},
  number   = {3},
  pages    = {489--506},
  volume   = {56},
  journal  = {IEEE Trans. Ultras. Ferroel. Freq. Cont.},
  month    = mar,
  pmid     = {19411209},
  year     = {2009},
}

@Article{mallart_adaptive_1994,
  author     = {Mallart, Raoul and Fink, Mathias},
  title      = {Adaptive focusing in scattering media through sound‐speed inhomogeneities: {The} van {Cittert} {Zernike} approach and focusing criterion},
  doi        = {10.1121/1.410562},
  issn       = {0001-4966},
  number     = {6},
  pages      = {3721--3732},
  url        = {https://asa.scitation.org/doi/10.1121/1.410562},
  urldate    = {2022-04-04},
  volume     = {96},
  journal    = {J. Acoust. Soc. Am.},
  month      = dec,
  shorttitle = {Adaptive focusing in scattering media through sound‐speed inhomogeneities},
  year       = {1994},
}

@Article{chau_locally_2019,
  author   = {Chau, Gustavo and Jakovljevic, Marko and Lavarello, Roberto and Dahl, Jeremy},
  title    = {A {Locally} {Adaptive} {Phase} {Aberration} {Correction} ({LAPAC}) {Method} for {Synthetic} {Aperture} {Sequences}},
  doi      = {10.1177/0161734618796556},
  issn     = {1096-0910},
  number   = {1},
  pages    = {3--16},
  volume   = {41},
  journal  = {Ultrason. Imag.},
  month    = jan,
  pmid     = {30222052},
  year     = {2019},
}

@Article{cho_efficient_2009,
  author   = {Cho, M. H. and Kang, L. H. and Kim, J. S. and Lee, S. Y.},
  title    = {An efficient sound speed estimation method to enhance image resolution in ultrasound imaging},
  doi      = {10.1016/j.ultras.2009.06.005},
  issn     = {1874-9968},
  number   = {8},
  pages    = {774--778},
  volume   = {49},
  journal  = {Ultrasonics},
  month    = dec,
  pmid     = {19635626},
  year     = {2009},
}

@Article{jaeger_full_2015,
  author   = {Jaeger, Michael and Robinson, Elise and Akarçay, H Günhan and Frenz, Martin},
  title    = {Full correction for spatially distributed speed-of-sound in echo ultrasound based on measuring aberration delays via transmit beam steering},
  doi      = {10.1088/0031-9155/60/11/4497},
  issn     = {0031-9155, 1361-6560},
  number   = {11},
  pages    = {4497--4515},
  url      = {https://iopscience.iop.org/article/10.1088/0031-9155/60/11/4497},
  urldate  = {2021-11-02},
  volume   = {60},
  journal  = {Phys. Med. Biol.},
  month    = jun,
  year     = {2015},
}

@Article{bendjador_svd_2020,
  author     = {Bendjador, Hanna and Deffieux, Thomas and Tanter, Mickaël},
  title      = {The {SVD} {Beamformer}: {Physical} {Principles} and {Application} to {Ultrafast} {Adaptive} {Ultrasound}},
  doi        = {10.1109/TMI.2020.2986830},
  issn       = {1558-254X},
  number     = {10},
  pages      = {3100--3112},
  volume     = {39},
  journal    = {IEEE Trans. Med. Imag.},
  month      = oct,
  shorttitle = {The {SVD} {Beamformer}},
  year       = {2020},
}

@Article{labeyrie_attainment_1970,
  author   = {Labeyrie, A.},
  title    = {Attainment of {Diffraction} {Limited} {Resolution} in {Large} {Telescopes} by {Fourier} {Analysing} {Speckle} {Patterns} in {Star} {Images}},
  issn     = {0004-6361},
  pages    = {85},
  url      = {https://ui.adsabs.harvard.edu/abs/1970A&A.....6...85L/abstract},
  urldate  = {2022-04-12},
  volume   = {6},
  journal  = {Astronomy and Astrophysics},
  month    = may,
  year     = {1970},
}

@Article{lambert_distortion_2020,
  author   = {Lambert, William and Cobus, Laura A. and Frappart, Thomas and Fink, Mathias and Aubry, Alexandre},
  title    = {Distortion matrix approach for ultrasound imaging of random scattering media},
  doi      = {10.1073/pnas.1921533117},
  issn     = {0027-8424, 1091-6490},
  number   = {26},
  pages    = {14645--14656},
  url      = {http://www.pnas.org/lookup/doi/10.1073/pnas.1921533117},
  urldate  = {2021-12-23},
  volume   = {117},
  journal  = {Proc. Nat. Acad. Sci. USA},
  month    = jun,
  year     = {2020},
}

@Article{nock_phase_1989,
  author   = {Nock, L. and Trahey, G. E. and Smith, S. W.},
  title    = {Phase aberration correction in medical ultrasound using speckle brightness as a quality factor},
  doi      = {10.1121/1.397889},
  issn     = {0001-4966},
  number   = {5},
  pages    = {1819--1833},
  volume   = {85},
  journal  = {J. Acoust. Soc. Am.},
  month    = may,
  pmid     = {2732378},
  year     = {1989},
}

@Article{ali_local_2021,
  author   = {Ali, Rehman and Telichko, Arsenii V. and Wang, Huaijun and Sukumar, Uday K. and Vilches-Moure, Jose G. and Paulmurugan, Ramasamy and Dahl, Jeremy J.},
  title    = {Local {Sound} {Speed} {Estimation} for {Pulse}-{Echo} {Ultrasound} in {Layered} {Media}},
  doi      = {10.1109/TUFFC.2021.3124479},
  issn     = {0885-3010, 1525-8955},
  pages    = {500--511},
  url      = {https://ieeexplore.ieee.org/document/9597614/},
  urldate  = {2021-12-08},
  volume   = {69},
  journal  = {IEEE Trans. Ultras. Ferroel. Freq. Cont.},
  year     = {2022},
}

@Article{Staehli2023,
  author    = {Stähli, Patrick and Becchetti, Chiara and Korta Martiartu, Naiara and Berzigotti, Annalisa and Frenz, Martin and Jaeger, Michael},
  title     = {First-in-human diagnostic study of hepatic steatosis with computed ultrasound tomography in echo mode},
  doi       = {10.1038/s43856-023-00409-3},
  issn      = {2730-664X},
  number    = {1},
  pages     = {176},
  volume    = {3},
  journal   = {Commun. Med.},
  month     = dec,
  publisher = {Springer Science and Business Media LLC},
  year      = {2023},
}

@InBook{Simson2023,
  author    = {Simson, Walter and Zhuang, Louise and Sanabria, Sergio J. and Antil, Neha and Dahl, Jeremy J. and Hyun, Dongwoon},
  booktitle = {Medical Image Computing and Computer Assisted Intervention – MICCAI 2023},
  title     = {Differentiable Beamforming for Ultrasound Autofocusing},
  doi       = {10.1007/978-3-031-43999-5_41},
  isbn      = {9783031439995},
  pages     = {428--437},
  publisher = {Springer Nature Switzerland},
  issn      = {1611-3349},
  year      = {2023},
}

@Article{Bazulin2022,
  author    = {Bazulin, Evgeny and Goncharsky, Alexander and Romanov, Sergey and Seryozhnikov, Sergey},
  title     = {Ultrasound transmission and reflection tomography for nondestructive testing using experimental data},
  doi       = {10.1016/j.ultras.2022.106765},
  issn      = {0041-624X},
  pages     = {106765},
  volume    = {124},
  journal   = {Ultrasonics},
  month     = aug,
  publisher = {Elsevier BV},
  year      = {2022},
}

@Article{Shiraishi2023,
  author    = {Shiraishi, Kazuya and Watanabe, Toshiki},
  title     = {Seismic reflection imaging of deep crustal structures using local earthquakes in the {K}anto region, {J}apan},
  doi       = {10.1186/s40623-023-01772-0},
  issn      = {1880-5981},
  number    = {1},
  pages     = {14},
  volume    = {75},
  journal   = {EPS},
  month     = jan,
  publisher = {Springer Science and Business Media LLC},
  year      = {2023},
}

@Article{Simson2024,
  author       = {Simson, W. A. and Paschali, M. and Sideri-Lampretsa, V. and Navab, N. and Dahl, J. J.},
  date         = {2024},
  journaltitle = {Ultrasonics},
  title        = {Investigating pulse-echo sound speed estimation in breast ultrasound with deep learning},
  doi          = {10.1016/j.ultras.2023.107179},
  pages        = {107179},
  volume       = {137},
  journal      = {Ultrasonics},
  year         = {2024},
}

@Article{Gordon1995,
  author    = {Gordon, Paula B. and Goldenberg, S. Larry},
  date      = {1995},
  title     = {Malignant breast masses detected only by ultrasound. A retrospective review},
  issn      = {1097-0142},
  number    = {4},
  pages     = {626--630},
  volume    = {76},
  journal   = {Cancer},
  publisher = {Wiley},
  year      = {1995},
}

@Article{lambert_reflection_2020,
  author    = {Lambert, William and Cobus, Laura A. and Couade, Mathieu and Fink, Mathias and Aubry, Alexandre},
  title     = {Reflection Matrix Approach for Quantitative Imaging of Scattering Media},
  doi       = {10.1103/physrevx.10.021048},
  issn      = {2160-3308},
  number    = {2},
  pages     = {021048},
  volume    = {10},
  journal   = {Phys. Rev. X},
  month     = jun,
  publisher = {American Physical Society (APS)},
  year      = {2020},
}

@Article{lambert_ultrasound_2022,
  author    = {Lambert, William and Cobus, Laura A. and Robin, Justine and Fink, Mathias and Aubry, Alexandre},
  title     = {Ultrasound Matrix Imaging{—Part II: T}he Distortion Matrix for Aberration Correction Over Multiple Isoplanatic Patches},
  doi       = {10.1109/tmi.2022.3199483},
  issn      = {1558-254X},
  number    = {12},
  pages     = {3921--3938},
  volume    = {41},
  journal   = {IEEE Trans. Med. Imag.},
  month     = dec,
  publisher = {Institute of Electrical and Electronics Engineers (IEEE)},
  year      = {2022},
}

@Article{lambert_ultrasound_2022a,
  author    = {Lambert, William and Robin, Justine and Cobus, Laura A. and Fink, Mathias and Aubry, Alexandre},
  title     = {Ultrasound Matrix Imaging{—Part I: T}he Focused Reflection Matrix, the F-Factor and the Role of Multiple Scattering},
  doi       = {10.1109/tmi.2022.3199498},
  issn      = {1558-254X},
  number    = {12},
  pages     = {3907--3920},
  volume    = {41},
  journal   = {IEEE Trans. Med. Imag.},
  month     = dec,
  publisher = {Institute of Electrical and Electronics Engineers (IEEE)},
  year      = {2022},
}

@Article{Almeida2008,
  author    = {Almeida, Alessandro-Moura},
  title     = {Fatty liver disease in severe obese patients: Diagnostic value of abdominal ultrasound},
  doi       = {10.3748/wjg.14.1415},
  issn      = {1007-9327},
  number    = {9},
  pages     = {1415},
  volume    = {14},
  journal   = {World J. Gastroenterol.},
  publisher = {Baishideng Publishing Group Inc.},
  year      = {2008},
}

@Article{Hinkelman1998,
  author    = {Hinkelman, Laura M. and Mast, T. Douglas and Metlay, Leon A. and Waag, Robert C.},
  title     = {The effect of abdominal wall morphology on ultrasonic pulse distortion. Part I. Measurements},
  doi       = {10.1121/1.423946},
  issn      = {1520-8524},
  number    = {6},
  pages     = {3635--3649},
  volume    = {104},
  journal   = {J. Acoust. Soc. Am.},
  month     = dec,
  publisher = {Acoustical Society of America (ASA)},
  year      = {1998},
}

@Article{Browne2005,
  author    = {Browne, Jacinta E. and Watson, Amanda J. and Hoskins, Peter R. and Elliott, Alex T.},
  title     = {Investigation of the effect of subcutaneous fat on image quality performance of 2D conventional imaging and tissue harmonic imaging},
  doi       = {10.1016/j.ultrasmedbio.2005.03.012},
  issn      = {0301-5629},
  number    = {7},
  pages     = {957--964},
  volume    = {31},
  journal   = {Ultrasound Med. Biol.},
  month     = jul,
  publisher = {Elsevier BV},
  year      = {2005},
}

@Article{Lediju2009,
  author    = {Lediju, M.A. and Pihl, M.J. and Hsu, S.J. and Dahl, J.J. and Gallippi, C.M. and Trahey, G.E.},
  title     = {A motion-based approach to abdominal clutter reduction},
  doi       = {10.1109/tuffc.2009.1331},
  issn      = {0885-3010},
  number    = {11},
  pages     = {2437--2449},
  volume    = {56},
  journal   = {IEEE Trans. Ultrason. Ferroel. Freq. Cont.},
  month     = nov,
  publisher = {Institute of Electrical and Electronics Engineers (IEEE)},
  year      = {2009},
}

@Article{Scorza2015,
  author    = {Scorza, A and Conforto, S and D’Anna, C and Sciuto, S.A},
  title     = {A Comparative Study on the Influence of Probe Placement on Quality Assurance Measurements in B-mode Ultrasound by Means of Ultrasound Phantoms},
  doi       = {10.2174/1874120701509010164},
  issn      = {1874-1207},
  number    = {1},
  pages     = {164--178},
  volume    = {9},
  journal   = {Open Biomed. Eng. J.},
  month     = jul,
  publisher = {Bentham Science Publishers Ltd.},
  year      = {2015},
}

@Article{Nicolaides1994,
  author    = {Nicolaides, K. H. and Brizot, M. L. and Snijders, R. J. M.},
  title     = {Fetal nuchal translucency: ultrasound screening for fetal trisomy in the first trimester of pregnancy},
  doi       = {10.1111/j.1471-0528.1994.tb11946.x},
  issn      = {1471-0528},
  number    = {9},
  pages     = {782--786},
  volume    = {101},
  journal   = {BJOG},
  month     = sep,
  publisher = {Wiley},
  year      = {1994},
}

@Article{Snijders1998,
  author    = {Snijders, RJM and Noble, P and Sebire, N and Souka, A and Nicolaides, KH},
  title     = {{UK} multicentre project on assessment of risk of trisomy 21 by maternal age and fetal nuchal-translucency thickness at 10–14 weeks of gestation},
  doi       = {10.1016/s0140-6736(97)11280-6},
  issn      = {0140-6736},
  number    = {9125},
  pages     = {343--346},
  volume    = {352},
  journal   = {The Lancet},
  month     = aug,
  publisher = {Elsevier BV},
  year      = {1998},
}

@Article{balondrade_multi-spectral_2023,
  author       = {Balondrade, Paul and Barolle, Victor and Guigui, Nicolas and Auriant, Emeric and Rougier, Nathan and Boccara, Claude and Fink, Mathias and Aubry, Alexandre},
  date         = {2024},
  journaltitle = {Nature Photonics},
  title        = {Multi-{Spectral} {Reflection} {Matrix} for {Ultra}-{Fast} {3D} {Label}-{Free} {Microscopy}},
  pages        = {1097-1104},
  volume       = {18},
  journal      = {Nat. Photon.},
  year         = {2024},
}

@Article{Ahmed2024,
  author    = {Ahmed, Rifat and Trahey, Gregg},
  title     = {Spatial Ambiguity Correction in Coherence-based Average Sound Speed Estimation},
  doi       = {10.1109/tuffc.2024.3440832},
  issn      = {1525-8955},
  pages     = {1244--1254},
  volume    = {71},
  journal   = {IEEE Trans. Ultrason. Ferroel. Freq. Cont.},
  publisher = {Institute of Electrical and Electronics Engineers (IEEE)},
  year      = {2024},
}

@Article{giraudat_unveiling_2023,
  author    = {Giraudat, Elsa and Burtin, Arnaud and Ber, Arthur Le and Fink, Mathias and Komorowski, Jean-Christophe and Aubry, Alexandre},
  title     = {Matrix imaging as a tool for high-resolution monitoring of deep volcanic plumbing systems with seismic noise},
  doi       = {10.1038/s41566-024-01479-y},
  pages     = {509},
  volume    = {5},
  copyright = {Creative Commons Attribution 4.0 International},
  journal   = {Commun. Earth Environ.},
  year      = {2024},
}

@Misc{Vraalstad2024,
  author    = {Vrålstad, Anders Emil and Fosodeder, Peter and Deibele, Karin Ulrike and Nyrnes, Siri Ann and Rindal, Ole Marius Hoel and Skoura-Torvik, Vibeke and Mienkina, Martin and Måsøy, Svein-Erik},
  title     = {Coherence Based Sound Speed Aberration Correction -- with clinical validation in fetal ultrasound},
  doi       = {10.48550/ARXIV.2411.16551},
  note      = {arXiv:2411.16551},
  copyright = {Creative Commons Attribution Non Commercial No Derivatives 4.0 International},
  publisher = {arXiv},
  year      = {2024},
}

@Thesis{Bureau,
  author      = {Bureau, F.},
  institution = {Université Paris Sciences & Lettres},
  title       = {Multi-dimensional analysis of the reflection matrix for quantitative ultrasound imaging},
  type        = {PhD Thesis},
  note        = {PhD Thesis, PSL University},
  year        = {2023},
}

@Misc{Garnier2025,
  author    = {Garnier, Josselin and Giovangigli, Laure and Goepfert, Quentin and Millien, Pierre},
  title     = {Probing the speckle to estimate the effective speed of sound, a first step towards quantitative ultrasound imaging},
  doi       = {10.48550/ARXIV.2505.07566},
  note      = {arXiv:2505.07566},
  copyright = {Creative Commons Attribution 4.0 International},
  publisher = {arXiv},
  year      = {2025},
}

@Article{Giraudat2025,
  author  = {Giraudat, E. and Bureau, F. and Lambert, W. and Fink, M. and Aubry, A.},
  title   = {Self-Portrait of the Focusing Process in Speckle: {I.} Spatio-Temporal Imaging of Wave Packets in Complex Media},
  journal = {arXiv},
  year    = {2026},
}

\end{document}